\documentclass{aa}

\usepackage{subfig}
\usepackage{soul}       
\usepackage{xcolor}

\begin{document}

   \title{A ring-shaped starburst as a \\ galactic wind-generating mechanism}

  \subtitle{Morphology, emission, and mass ejection}

   \author{J.A. Osorio-Caballero \inst{1}, A. Rodr\'iguez-Gonz\'alez\inst{1} , Z. Meliani\inst{2} 
          }

   \institute{$^{1}$Instituto de Ciencias Nucleares, Universidad Nacional Aut\'onoma de M\'exico, Ap. 70-543, 04510 CDMX, M\'exico \\
   $^{2}$LUX, Observatoire de Paris, Université PSL, Sorbonne Université, CNRS, 92190 Meudon, France\\
 }

   \date{Received January 15, 2025; accepted March 16, 2025}

 
  \abstract
   {Star formation bursts promote the ejection of material from the hosting galaxies due to the momentum and energy injected by winds from massive stars and supernova explosions. Numerical or analytical models generally consider that the mass, momentum, and energy injections result from  bursts in a nuclear star formation region. However, star formation bursts have recently been observed in ring-like regions in the nuclear part of the galaxies. One example is NGC 253, which has shown a central toroidal burst and an asymmetric galactic wind observed in thermal X-ray emission. }
   {The general aim of this work is to study the effect of mechanical energy injection from stellar winds and supernova explosions in star-forming bursts distributed in rings around the nucleus of the galaxy NGC 253. Additionally, these partial objectives  allow us to analyse the asymmetry of the outflows due to the burst's position as well as to study the formation of filaments with optical emission and make comparisons with recent observations of galaxies with these types of star-forming bursts.}
   {We used the hydrodynamic code AMRVAC to simulate galactic wind ejection coming from a central ring-like starburst located at different vertical positions: at the centre and in 10 pc and 0.1 kpc from the centre. Our models considered the starburst evolution following the stellar population of an instantaneous and continuous burst, using the SB99 synthesis models, where the mechanical energy injected via SN and stellar wind are considered. To compare our results with the X-ray emission from the outflow from NGC253, we  built the H${\alpha}$ and X-ray emissions maps and quantified the mass flux of the galactic wind.}
   {We showed that including a ring-shaped starburst (RSS) generates a more complex structured wind than what would be expected for a spherical starburst injection. Besides the interaction between the wind generated by the RSS and the host galaxy, it can generate dense filamentary structures with H$\alpha$ emission. The mass flux analysis of our models shows that the variation in the vertical position of the starburst can generate a variation in the mass flux of each lobe of the wind up to an order of magnitude. However, this difference is sustained only for a short period, with the flux tending to be symmetrical once it enters into a free-wind solution.}
   {The inclusion of RSS in our simulations generates a wind with a more defined structure, with a more collimated cone of injection. This allows it to break through the thick disk and generate a double flow galactic wind with an off-centre starburst, resulting in expected structures such as H${\alpha}$ filaments.    
   Through the change in the vertical position of the starburst about the host galaxy, our simulations were able to generate an asymmetric wind, revealing important differences in terms of size and mass flux. On the gas multi-phase structure of each side of the wind, the level of variation is in direct correlation with the variation in position, offering a plausible explanation for the kind of winds observed on NGC~253.}

\titlerunning{A ring-shaped starburst as a galactic wind-generating mechanism}
\authorrunning{Osorio-Caballero et al.} 
   \maketitle

\section{Introduction}

Galactic winds (GWs) have been associated with starburst events  \citep{Weedman_etal_1981ApJ...248..105W, Chevalier1985}. These powerful outflows are caused by the material ejected by the stars generated on set starburst; therefore, their mass flux rates and velocities are strongly correlated with the host galaxy's SFR \citep{Rubin_etal_2014ApJ...794..156R}. Furthermore, starburst galaxies are often called super wind galaxies due to the intensity of these outflows \citep{Thompson_Heckman_2024ARA&A..62..529T} and these super winds significantly influence the galactic disk, halo, and intergalactic medium.

Within galaxies, starburst events are typically concentrated in collections of super stellar clusters (SSCs) with masses exceeding 10~M$_{\odot}$ \citep{Portegies_Zwart_etal2010ARA&A..48..431P}, and compact at \(2-3~ \mathrm{pc}\) \citep{Ryon_etal_2017ApJ...841...92R}. These SSCs are predominantly located in the central regions of galaxies \citep{Holtzman1992, Melo2005, Li_etal_2015ApJS..216....6L}, where they play a crucial role in driving galactic winds \citep{zhang2018}. SSCs have  a large fraction of massive stars (\(M > 10 \, M_{\odot}\)), such as those observed in the  M82 galaxy \citep{Levy_etal_2024ApJ...973L..55L}. These massive stars inject mass and energy into the interstellar medium (ISM) through stellar winds during their lives and through supernovae at the end of their lives. These flows coalesce into a single wind emanating from each SSC. The combined contribution of winds from multiple SSCs in the central region of the Galaxy ultimately drives the galactic wind \citep[e.g.][]{Veilleux_etal_2005ARA&A..43..769V}.  
The formation and propagation of this wind also depend on the proprieties of the medium, such as the density and pressure profile of the host galaxy. These properties provide the appropriate structural conditions, such as a Laval nozzle, to channel and sustain the outflow \citep{ary2008,Rubin_etal_2014ApJ...794..156R}.

The wind generated by SSCs plays a significant role in the loss of gas, metals, and energy from the galactic centre \citep{Rob-Val2017} and in the enrichment of both the galactic medium \citep{DErcole1999, MacLow1999} and the intergalactic medium \citep{Sanders_etal_2023ApJ...942...24S}. It has been shown in previous studies that galactic winds, driven by
the SSCs winds, provide critical feedback on starburst activity within the
galaxy via observations \citep[e.g. ][]{Garnett_2002ApJ...581.1019G, Hirschauer_etal_2018AJ....155...82H, zhang2018, Krieger2021} and numerical simulations \citep{Bertone2007, ary2008, RodriguezRagaCanto2009, Rodriguez-gonzalez2011, Rob-Val2017}. This feedback alters the chemical evolution of the galaxy and across longer timescales, it also affects  star formation in the host galaxies \citep{Bertone2007}.

Observations show that SSCs are mainly distributed within the nuclear starburst region, which is the closest region of the galactic disk, as in the case of the archetypal starburst M82 (catalog built with HST observations \citep{Mayya_etal_2008ApJ...679..404M}, and shown in other recent observations with JWST \citep{Levy_etal_2024ApJ...973L..55L}).  In some starburst galaxies and galaxies with active galactic nuclei (AGNs), a fraction of SSCs are observed in nuclear rings \citep{Knapen_2005A&A...429..141K, Mazzuca_etal_2008ApJS..174..337M}. Examples include the galaxy NGC~253 \citep{Levy2022} and the galaxy NGC~7552 \citep{Brandl_etal_2012A&A...543A..61B}. NGC~253 is a starburst spiral galaxy located at a distance of 3.5~Mpc \cite{Rekola2005}. It exhibits a distribution of SSCs within an elliptical (ring) starburst located in the galactic centre. The ellipse has a semi-major axis of
$\sim$ 110 pc, a semi-minor axis of $\sim$60 pc, and the SSC inside the ellipse has an orbital period of
period of $\approx$ 6 Myr \citep{Levy2022} as well as an SFR of 2~M$_{\odot}$~yr$^{-1}$ \citep{Leroy2015}. 
The galactic wind in NGC253 is observed in H$\alpha$ up to 2~kpc and in X-rays up to 9~kpc from the galactic plane \citep{Strickland_2002}. The mass flux rate is estimated to be 2.8~M$_{\odot}$~yr$^{-1}$ for the northern outflow and 3.2~M$_{\odot}$~yr$^{-1}$ for the southern \citep{Lopez2023}, which corresponds to a 12.5$\%$ difference between the north-south flows.

Galactic winds exhibit complex dynamics with different velocities and fluxes at different galactic latitudes and altitudes. They also show an interstellar multi-phase structure, including very hot, hot, warm, cold, and relativistic (cosmic ray) phases, each characterised by different chemical states \citep{Heckman_Thompson_2017hsn..book.2431H, Thompson_Heckman_2024ARA&A..62..529T}. Such properties have been observed in the starburst galaxy M82 \citep{zhang2018, Krieger2021,Lopez2023}. Recently, it was suggested that the transverse stratification of galactic winds, characterised by hot and cold components, could arise from the interaction of isotropic winds from central galactic superstar clusters with the galactic disk \citep{Strickland-Stevens-2000, Strickland_2002, Melo2005, Meliani2024}. In this scenario, the wind gets shocked, and deflected by the galactic disk, then becomes denser, decreasing the temperature under the influence of radiative cooling. An alternative scenario proposes that the stratification results from the combined outflows of SSCs within a ring structure \citep{Nguyen2022}, where the winds are heated by terminal shocks formed when converging flows collide near the galactic axis; however, this study was performed on the basis of a uniform medium, which lacked any possible interaction of the wind with the galaxy disk.

This paper investigates how these two processes, the winds from SSCs within a ring structure and the interaction of the galactic wind with the galactic disk, contribute to the development of multi-phase and multi-dynamic galactic winds. In addition, we investigate the off-disk displacement of the ring and its asymmetry between the northern and southern regions. We aim to study how the wind can drag the disk material and how we can have a non-symmetric northern and southern flow such as the one observed in NGC~253 \citep{Lopez2023}.

NGC~253 galaxy  was specifically chosen as a case study since it has a well-characterised ring-shaped starburst (RSS), as well as a multi-frequency observed wind; this is in part thanks to the galaxy inclination of 76° \cite{Sanders2003}. Other galaxies also fulfil some of these characteristics, such as NGC~7552, which has a well-studied starburst ring \citep{Brandl_etal_2012A&A...543A..61B}; however, the inclination of this galaxy prevents us from determining the presence of a wind. There is also the example of NGC~7771, whose starburst ring has been well characterised by \cite{Smith1999Apj}, although it does  not show any evidence of an outflow in the galaxy.

This paper is structured as follows. Section~1 reviews previous studies of this type of event, focussing on the starburst galaxy NGC 253. In Section~2, we describe the factors we considered for our simulations, including the position and structure of the SSCs and the galaxy. Section~3 presents the results of our simulations and their analysis. Finally, in Section~4, we summarise the main results and conclusions of this work.

\section{Numerical models}

We performed axisymmetric hydrodynamical simulations in spherical coordinates, using axial boundary conditions; in particular, taking the axes perpendicular to the disk and passing through its centre. The wind (mass, momentum, and energy fluxes) are injected from a ring surrounding the galactic axis.
The hydrodynamical simulations of galactic winds were performed using the MPI-AMRVAC code \citep{Keppens2021}, which solves the conservation equations considering an optically thin radiative cooling by including it as a source term in the energy equation \citep{VanMarleKeppens_2011CF.....42...44V}. The governing equations are given by

\begin{eqnarray}\label{Eq} 
\frac{\partial \rho}{\partial t} + \vec{\nabla} \cdot (\rho \vec{v}) = 0 , \label{eq:euler1}\\ 
\frac{\partial \rho \vec{v}}{\partial t} + \vec{\nabla} \cdot \left(\rho \vec{v} \cdot \vec{v} \right) + \nabla \left(p \right) = 0, \label{eq:euler2} \\
\frac{\partial e}{\partial t} + \vec{\nabla} \cdot \left[\left(e + p \right) \vec{v} \right] = -\left(\frac{\rho}{1.4\,m_{h}}\right)^2 \Lambda(T), \label{eq:euler3} 
\\
\end{eqnarray}
where $\rho$ is the mass density, $\vec{v}$ is the velocity, and $p$ is the thermal pressure. The symbol $m_{h}$ denotes the mass of a hydrogen atom and $\Lambda(T)$ is the cooling function from the study by \citet{Schure_etal_2009A&A...508..751S}, calibrated for solar metallicity and dependent on the temperature t=$1.1 m_{h} p/ (\rho k_{\rm B})$. 

In our simulations, we imposed a temperature floor of $100\,\mathrm{K}$ to ensure numerical stability and to prevent excessively short cooling timescales that would otherwise require prohibitively small timesteps; this is particularly important since we are modelling large-scale flows up to kiloparsec (kpc) scales. This commonly adopted numerical strategy is fundamental in simulations where detailed molecular and dust cooling processes are not explicitly included. For the physical configurations, input parameters, and spatial resolution explored in this study, we find that the gas temperature in the wind–disc interaction region remains above approximately $400\,\mathrm{K}$. Thus, the imposed temperature floor cannot be reached by any significant component of the flow and it does not affect the thermal structure or the emergence of a multi-phase configuration. We note that at lower temperatures ($T \lesssim 100\,\mathrm{K}$), additional cooling mechanisms such as molecular and dust cooling may become relevant and could alter the dynamics of the interaction region \citep[e.g.][]{Chen_Oh_2024MNRAS.530.4032C}.

The total energy density, $e$, is given by
\begin{equation}
e=  p/(\gamma-1) + \rho v^2/2,
\end{equation}
while $\gamma = 5/3$ is the constant polytropic index. 

Naturally, there are some physical processes that have not been included in the current project. Gravitational forces are assumed to have a negligible influence on the overall wind dynamics, given the large spatial scales, the relatively short simulation timescales, and the high velocity of the injected wind. However, gravity could affect the evolution of slower, cold clouds that may form in the interaction region between the wind and the ambient medium.
Cosmic rays, while known to contribute to the launching and amplification of galactic outflows \citep{Sike2025ApJ}, are not accounted for in this work. In the presence of a strong starburst, such as the one modelled here, current studies suggest that their impact remains subdominant \citep{Romano2025arXiv}.
Magnetic fields have not been taken in to count either, although they have been shown to influence the expansion of the forward shock and enhance the resulting X-ray emission \citep{Meliani2024}; notably, this effect seems to be relevant for magnetic fields on the range of $10^2-10^4$~G, which is two orders of magnitude higher than the one detected in NGC~253 \citep{Beck1994}.
These additional processes could be relevant, however, they lie beyond the scope of this paper and will be considered in future work.

\subsection{Interstellar medium structure}

We  constructed the galaxy medium structure, considering a turbulent warm disk and a hot halo, both in hydrodynamic equilibrium, following the approach used in \cite{Meliani2024} and the framework of \cite{Strickland-Stevens-2000}. The galactic potential is represented by combining King's distribution for the halo and the \cite{Miyamoto_Nagai1975PASJ...27..533M} model for the disk. In fact, the disk has a mass of $2\times10^{9}\,\mathrm{M}_{\odot}$, a mean density of $n_{\rm disk,0}=20\,\mathrm{cm}^{-3}$, and a temperature of $T_{\rm disk}=6.7\times10^{4}\,\mathrm{K}$. The halo is set with a mass of $6\times10^{9}\,\mathrm{M}_{\odot}$ and  a mean density of $n_{\rm halo,0}=0.2\,\mathrm{cm}^{-3}$, with a core radios of R$= 350$~pc and a mean temperature of $T_{\rm halo}=6.9\times10^{6}\,\mathrm{K}$.

\defcitealias{Chevalier1985}{CC85}
\defcitealias{2000}{C2000}
\defcitealias{RG2007}{RG07}

\subsection{Starburst properties}
The evolution of the wind from massive, spherical clusters and its interaction with the galactic medium has been studied from both analytical and semi-analytical perspectives \citep{Chevalier1985, Canto2000, RG2007}. In our study, the galactic wind is assumed to be driven by a stellar population within a cluster that is eccentric to the galactic core. In this scenario, the galactic wind emerges from a ring surrounding the galactic axis (Fig.~\ref{fig:torus}), with the wind direction perpendicular to the surface of the ring shape.

\begin{figure}[!h]
    \centering
    \includegraphics[width=\columnwidth]{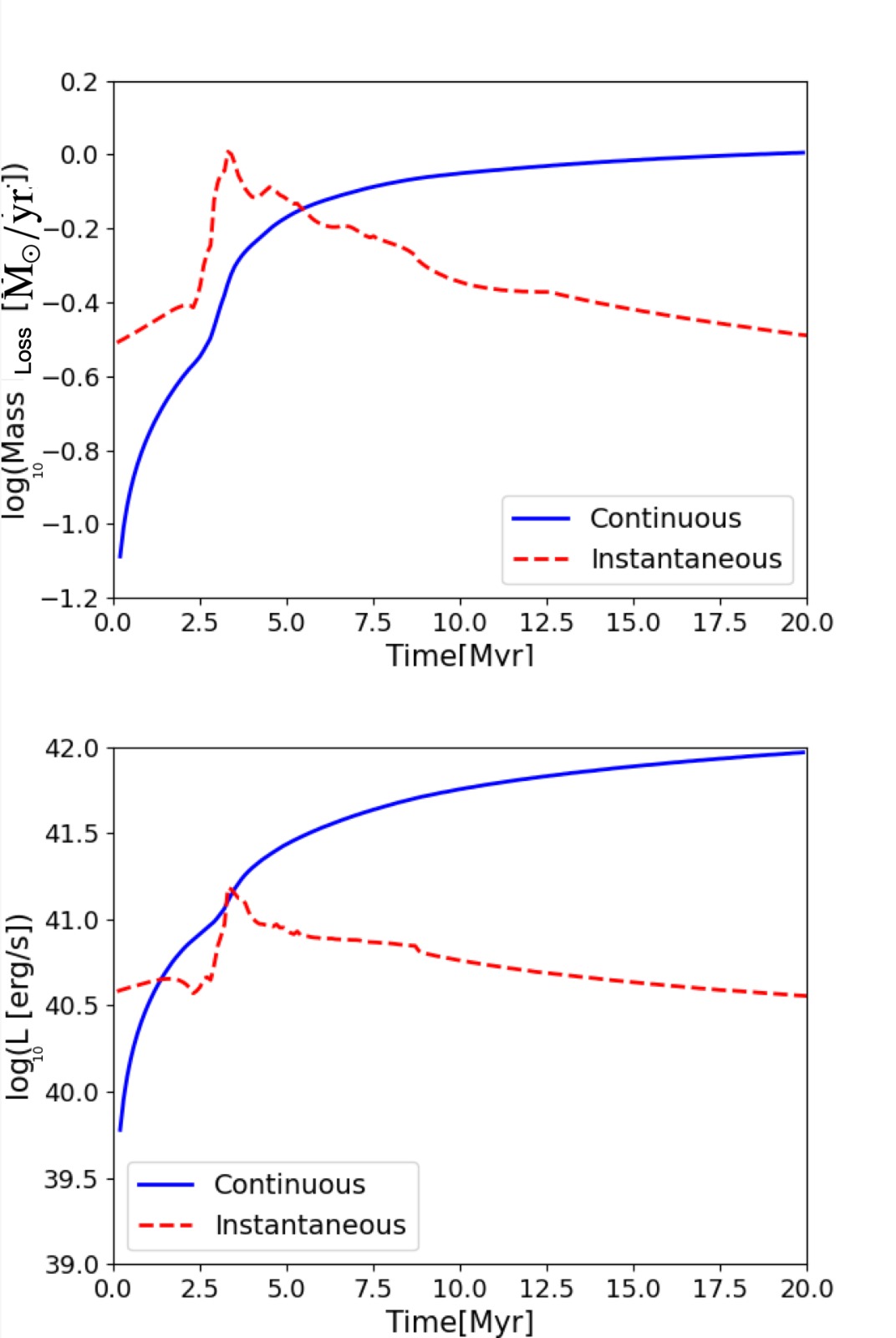}
    \caption[]{Mass loss (left side) and mechanical luminosity (right side) for an instantaneous burst (blue) and a continuous one (orange) obtained with SB99.}
    \label{fig:sb}
\end{figure}

The stellar population synthesis within the cluster was modeled according to two scenarios of star formation: continuous and instantaneous. The wind mass flux $\dot{M}(t)$ and mechanical luminosity $L_w(t)$ generated by the stellar cluster were derived using Starburst99 (SB99), assuming a Salpeter initial mass function \citep{Salpeter1955} with a stellar population that has masses ranging from 1 to 100~M$_\odot$. The resulting mass flux and mechanical luminosity as functions of time for both star formation scenarios are shown in Fig.~\ref{fig:sb}. The wind velocity is obtained as follows: $v_w^2(t) = \frac{L_w(t)}{\frac{1}{2} \dot{M}(t)}$.
We also considered the total mass of the cluster such that the maximum mass flux is 1~M$_\odot$/yr, consistent with the SFR of the star-forming ring of NGC 253 \citep{Leroy2015}.

\subsection{Initial setup}
 
To investigate the effect of star formation bursts with ring-like structures, we  performed hydrodynamical simulations based on a nuclear starburst with a ring-shaped distribution. The ring has a cross-sectional diameter of $R_e = 200$~pc and an inner radius of $R_i = 32.4$~pc. The nuclear starburst has a star formation rate of $2.5$~M$_\odot$/yr, corresponding to a higher mechanical luminosity of $2.47 \times 10^{41}$~erg/s.  

\begin{figure}[!h]
    \centering
    \includegraphics[width=0.5\columnwidth]{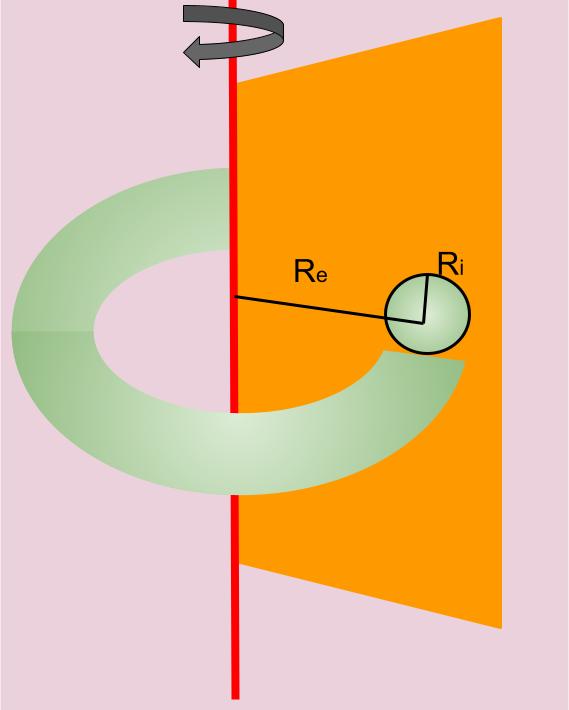}
    \caption[]{Set up for a ring-like starburst on an axisymmetric simulation, with a radius of injection equal to the internal radius of the torus, $R_i$, and positioned at a distance from the symmetry axes equal to the external radius of the torus: $R_e$ minus $R_i$.}
    \label{fig:torus}
\end{figure}

Using the evolutionary mass flux and mechanical luminosity presented in Fig.~\ref{fig:sb}, we carried out hydrodynamical simulations up to an evolutionary time of $10$~Myr; however, we compared our models at 4~Myr since this is the maximum time at which none of them have abandoned the simulation box. Figure~\ref{fig:torus} shows the ring-like star formation region from a 2D axisymmetric simulation perspective.

We have ran  five numerical models, varying the vertical altitude of the starburst ring along the galactic axis. These models incorporate the evolution of mass flux and mechanical luminosity for both continuous and instantaneous star formation scenarios. Table~\ref{tab:models} summarises the properties of the numerical models.

\begin{table}[]
    \centering
    \begin{tabular}{c c c}\hline\hline
         &Evolution type&Starburst vertical position  \\
         &&[kpc]\\\hline
         MCENc&Continuous&0 (centre)\\
         MOC10c&Continuous& 0.001 (off-centre)\\
         MOC100c&Continuous&0.1 (off-centre)\\
         MCENi&Instantaneous& 0 (centre)\\
         MOC100i&Instantaneous& 0.1 (off-centre)\\\hline
    \end{tabular}
    \caption{Specific proprieties for each model presented in the paper, indicating where the starburst position is measured on the positive vertical axis.}
    \label{tab:models}
\end{table}

\section{Results}

We present the analysis of our models and the results, focussing on the impact of the starburst ring's vertical position within the galactic disk.  
Our study examines the morphology, kinematics, and dynamics of the wind, its interaction with the galactic disk and halo, and the resulting emission.  
Additionally, we analysis of the mass flux at different altitudes in both the northern and southern hemispheres.

\subsection{Starburst in equatorial plane}

\begin{figure}[!h]
    \centering
    \includegraphics[width=\columnwidth]{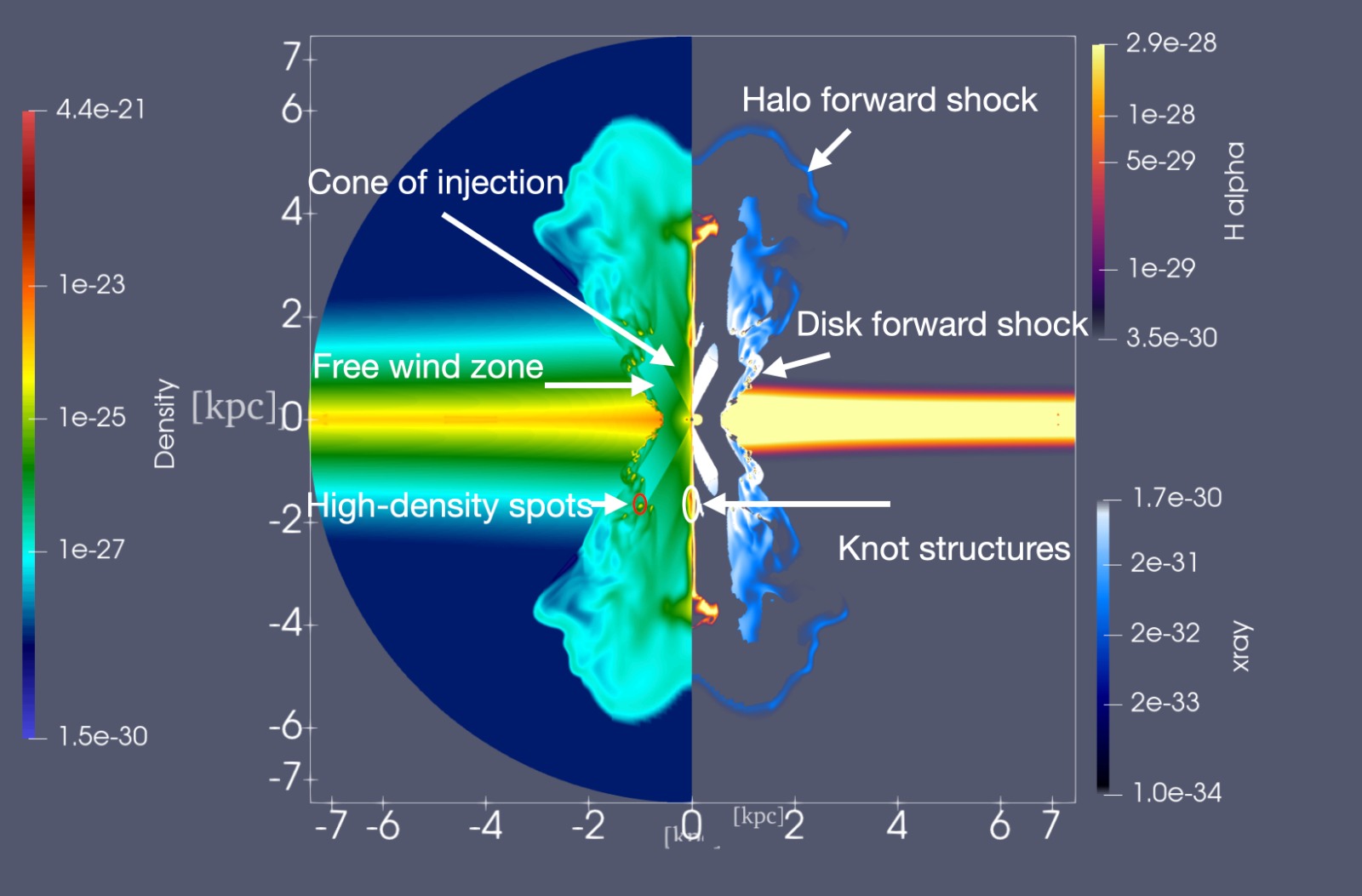}
    \caption[]{Simulation for a central and continuous starburst (MCENc) at t= 4 Myr; the left side shows density and the right side shows H$\alpha$(orange) and X-rays (blue) emissions.}
    \label{fig:centerSSCrho}
\end{figure}

We first considered the case of a starburst ring with a continuous burst located in the equatorial plane (MCENc). Figure~\ref{fig:centerSSCrho} (left) presents the density maps for the MCENc case, illustrating the main structures of the galactic bubble at t=$4$~Myr. The forward shock at high altitude, referred to as the halo forward shock, exhibits an arc-shaped structure within a region extending in altitude over $z = [4,6]$~kpc, with a radial expansion of $r \approx 2.5$~kpc. At lower altitudes $(z = [0,1.5]$~kpc), the disk material swept up by the wind escapes laterally, referred to as the disk forward shock, forming a dense shell that encloses the wind.

We also computed the H$\alpha$ and thermal soft X-ray emission maps, following the description in Appendix A. We note that we  considered the total thermal soft X-ray emission and we did not pay special attention to the contribution of charge exchange emission, which has been pointed out as an important contribution to the X-ray emission \citep{Zhang2014,Wu2020MNRAS.491.5621W}. To  consider this kind of emission, we would have had to track different gas species on the wind \citep{smith2014}, which would also allow us to generate synthetic lines of emission. This aspect will be addressed in future papers. However, according to \cite{Lopez2023}, in the case of NGC 253, this would imply an increase in the magnitude of the X-ray emission of 42\% in the central region of the galaxy and it
decreases to 20\% at 1~kpc away from it; therefore, at the scales of our simulation, the effect on the forward shock would be minimal.

Figure~\ref{fig:centerSSCrho} (right) presents these maps, where the H$\alpha$ emission is represented in purple-orange and the X-ray emission in blue bars. The X-ray emission originates from the hot, shocked, and deflected wind along the galactic axis within the conical structure. The case is similar in the central region of the simulation.
We have two main gas sources of H$\alpha$ emission: a) the first are the knot structures found in the centre of the wind, extending to 4~kpc in length and b) the second source can be found within a radius of 2~kpc, high-density clumps inside the secondary shock generate significant H$\alpha$ emission. This emission arises from the fragmentation of the dense, shocked shell formed by the galactic disk surrounding the free wind region. These structures may correspond to the filaments observed in M~82 \citep{ShopbelandBland1998}.

\subsection{Off-equatorial plane starburst}

\begin{figure}[!h]
    \centering
\includegraphics[width=0.7\columnwidth]{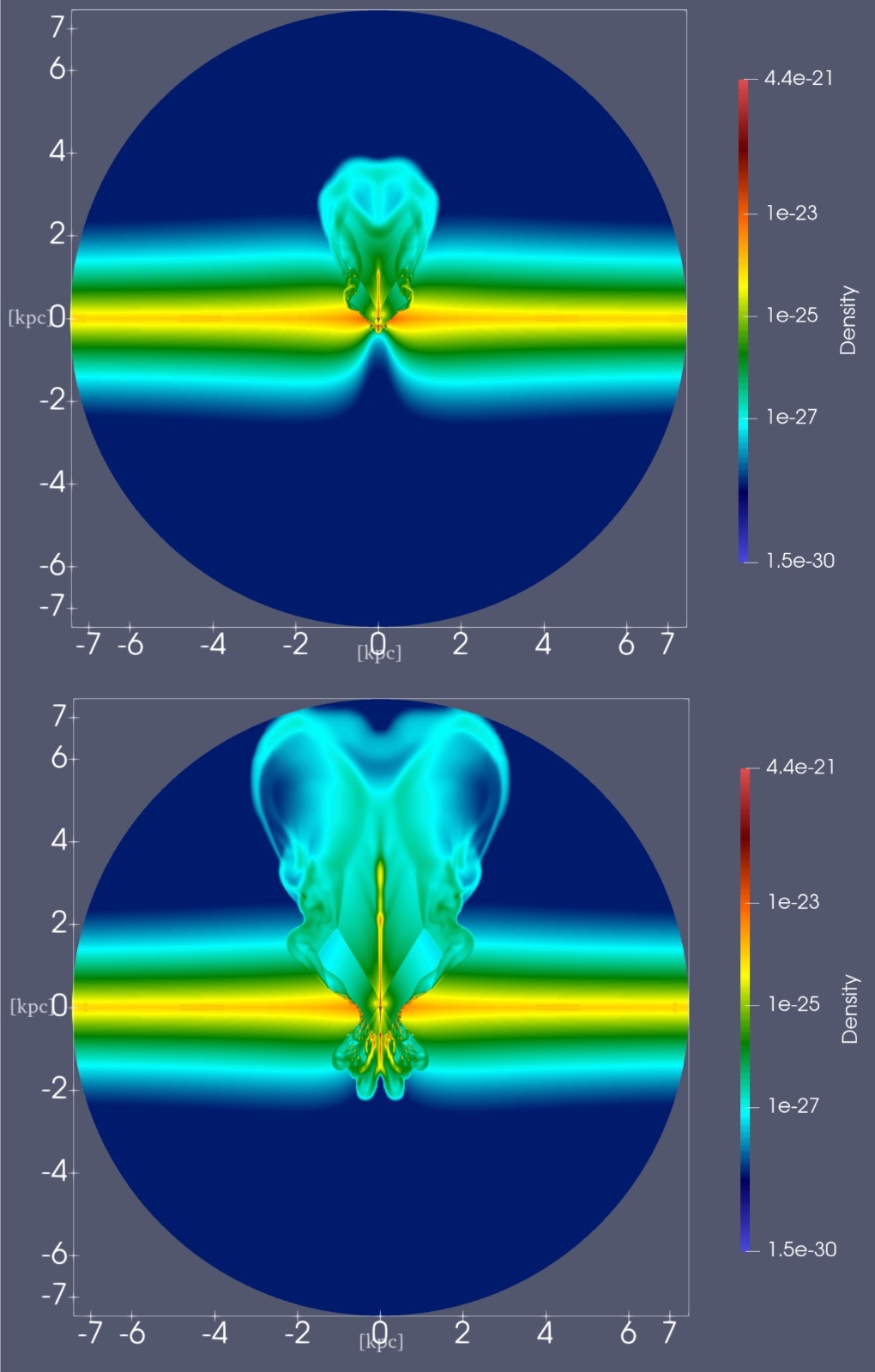}
    \caption[]{Gas density maps for the 0.1~kpc off-centre continuous model (MOC100c) going from a time of 2~Myr on the top to 4~Myr on the bottom.}
    \label{fig:OC2evolver}
\end{figure}

To reproduce the north–south asymmetry observed in galactic bubbles such as in NGC~253 \citep{Levy2022}, we performed a series of simulations in which the starburst ring was placed at various altitudes relative to the galactic equatorial plane. Both continuous and instantaneous wind injection scenarios were considered. Figure~\ref{fig:OC2evolver} displays the gas mass density for model MOC100c at t=$2$~Myr (top panel) and t=$4$~Myr (bottom panel). In this model, the starburst ring is located 100~pc above the galactic equatorial plane, on the northern side.

This asymmetric configuration results in a differential evolution of the galactic bubble between the northern and southern hemispheres. On the northern side, the galactic wind expands more rapidly through the disk, being closer to the low-density surface of the galaxy-disk. Conversely, on the southern side, the wind must propagate through the dense midplane of the galactic disk, entraining a larger amount of galactic-disk material. Moreover, the interaction with this denser region generates stronger shocks, which enhance radiative cooling losses. These combined effects significantly decelerate the expansion of the southern bubble compared to its northern counterpart.

A comparison between model MCENc, where the starburst cluster is located in the galactic equatorial plane (Fig.~\ref{fig:OC2evolver}), and model MOC100c, where the cluster is offset by 100~pc above the plane (Fig.~\ref{fig:centerSSCrho}), reveals morphological differences in the resulting galactic bubbles. On the northern side, the bubble in model MOC100c extends up to 1~kpc in height and is approximately 1~kpc wider than in model MCENc. Furthermore, the northern bubble in MOC100c exhibits a more laminar flow, with smoother structures and fewer dense filaments.

In contrast, on the southern side, the bubble in model MOC100c is significantly smaller, reaching only about 1~kpc in both height and width, and it remains confined within the galactic disk. In addition, it exhibits more prominent and denser filamentary structures, with a length greater than 1~kpc; this increment in length is due to the wind fracturing a larger part of the disk of the host galaxy. These substantial differences are induced by a vertical displacement of the starburst ring of only 100~pc.

\begin{figure}[!h]
    \centering
    \includegraphics[width=\columnwidth]{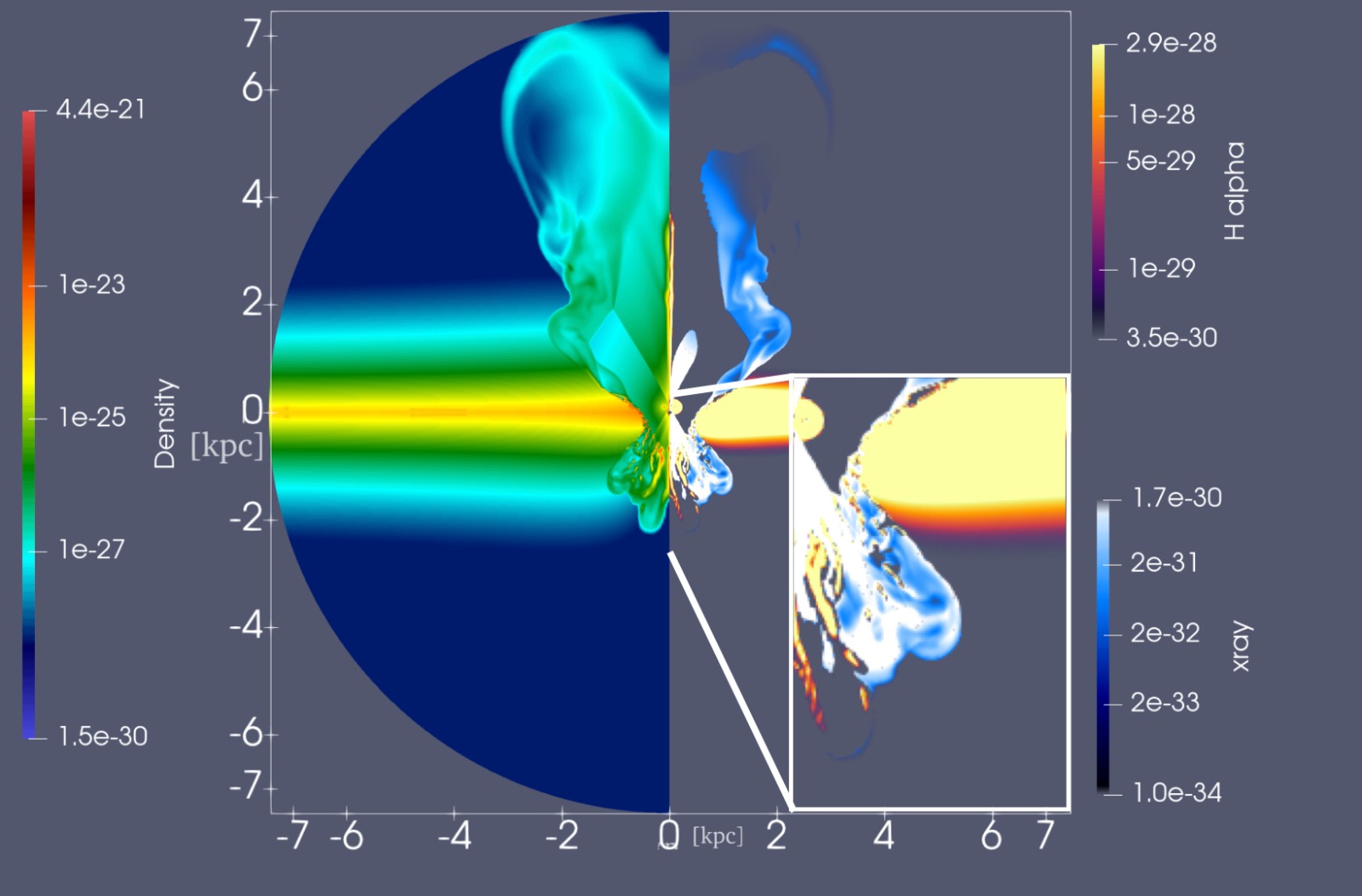}
    \caption[]{H$\alpha$ and X-ray emission maps of the wind simulated for the off-centre continuous starburst model (MOC100c), shifted by 0.1~kpc, at t=$4$~Myr.}
    \label{fig:OC2SSCHalpha}
\end{figure}

The northern and southern flows have a multi-phase structure. In the centre, a hot, conical wind arises from the deflection of the galactic wind along the galactic axis. This central wind extends to the terminal shock, reaching 4~kpc on the northern side, but only 0.25~kpc on the southern side. It is surrounded by a cooler, free-flowing galactic wind, which terminates at 2 kpc in the north and again at 0.25 kpc in the south. This is surrounded by a conical, cold shell formed by the interaction of the shocked, deflected galactic wind with the dense galactic disk. This shell also extends to 2 kpc in the north and 0.25 kpc in the south. Filaments form mainly on the southern side, where the shocks are stronger. On the northern side, filaments develop at the lateral edges of the cold shell and persist due to the mass and momentum deposited by the galactic wind. These structures continue to emit in H$\alpha$ (see Fig.~\ref{fig:OC2SSCHalpha}). In addition, on the northern side, filaments also emerge in the outer cold, dense shell surrounding the free-wind region, as well as behind the terminal shock.
In addition, high-density clumps appear near the centre at early times in this region, but quickly dissipate. These clumps are similar to the southern filaments, although they lack sufficient mass to grow to the same size as those seen in the MOC100c model.

The H$\alpha$ and X-ray emission maps shown in the right panel of Fig.~\ref{fig:OC2SSCHalpha} show that the southern flow has consistent X-ray emission throughout the wind. In contrast, the X-ray emission in the northern flow is mainly due to the swept-up shocked gas by the forward shock, resulting from the interaction between the galactic wind with the galactic halo gas and the galactic wind deflected by the galactic disk.

\begin{figure}[!h]
    \centering
    \includegraphics[width=\columnwidth]{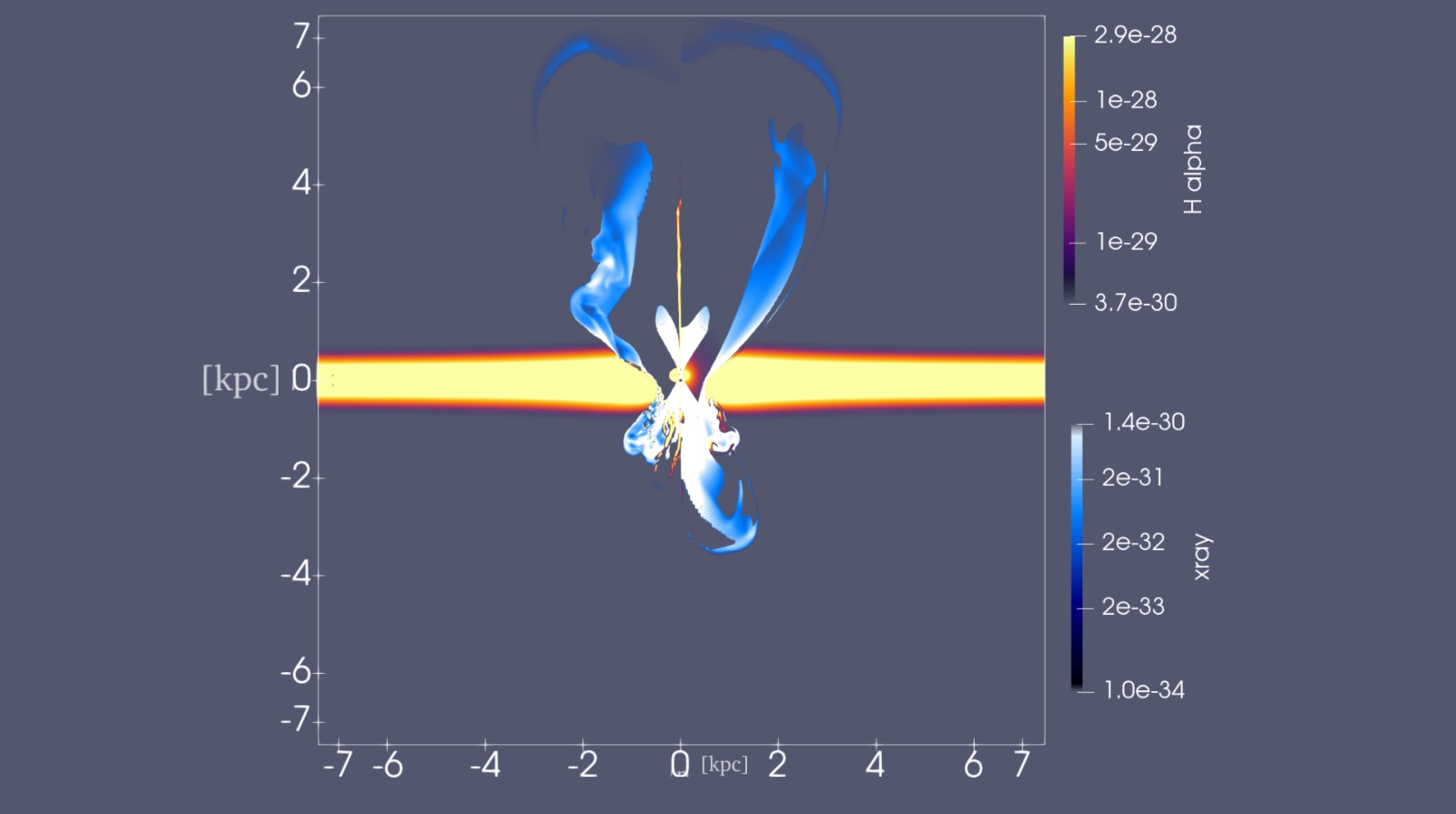}
    \caption[]{H$\alpha$ and X-ray emissions of the wind simulated for a 0.1~kpc off-centre continuous starburst (MOC100c)  at t=4~Myr Left:\ Case with cooling. Right:  Adiabatic evolution.}
    \label{fig:OC2SSCHalphaCNC}
\end{figure}

It is important to note that the formation of filaments is driven by disk fragmentation, which would not occur without including radiative cooling in our simulations, as highlighted by \cite{ary2008}. To demonstrate this effect, we performed an additional simulation of model MOC100c in which radiative cooling was switched off. The resulting radiative emission at t=$4$~Myr for the simulations with and without cooling is shown in Fig. \ref{fig:OC2SSCHalphaCNC} (left and right panels, respectively). In the absence of cooling, the galactic bubble expands more uniformly and develops significantly fewer filamentary structures.

\begin{figure}[!h]
    \centering
    \includegraphics[width=\columnwidth]{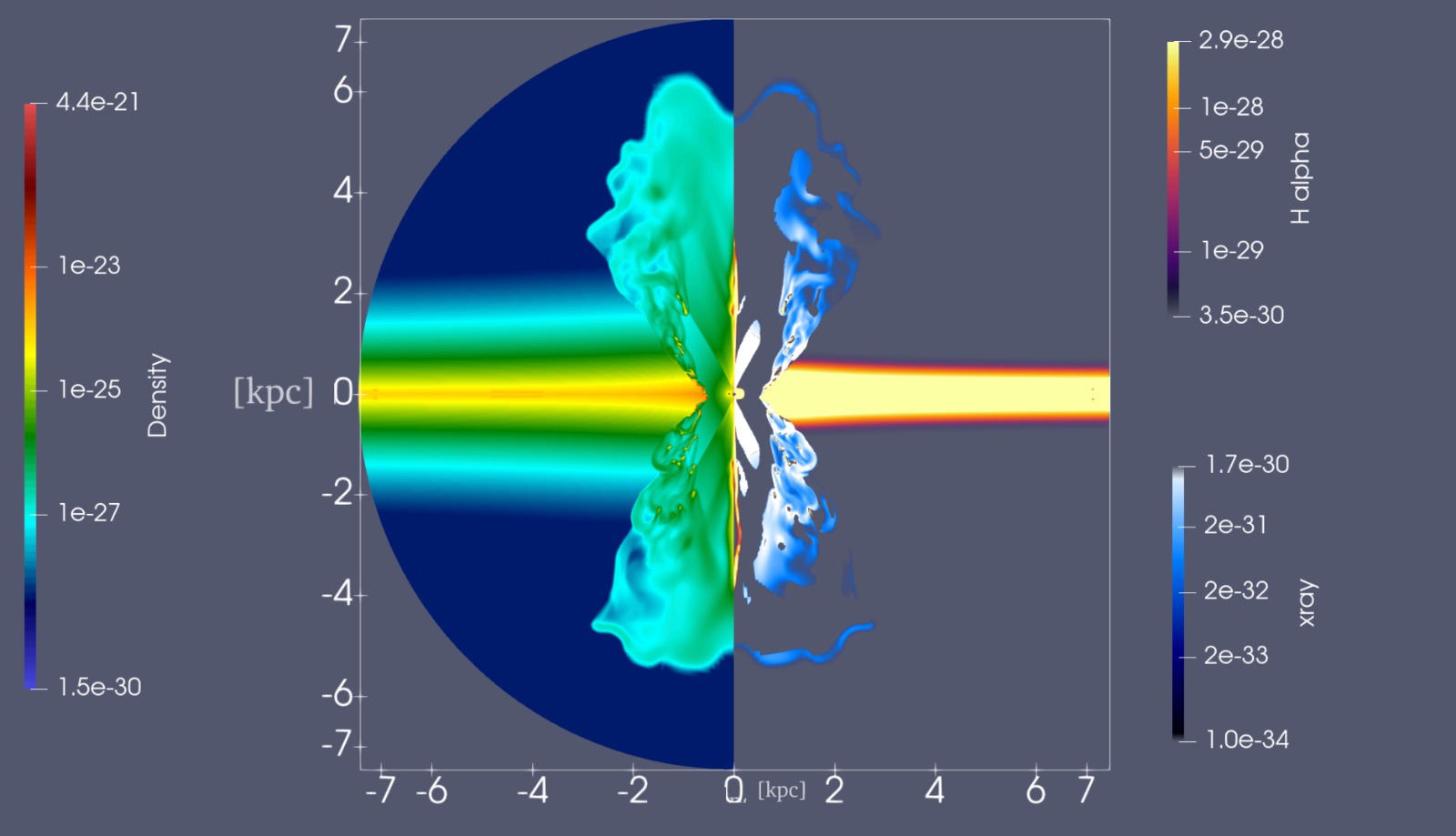}
    \caption[]{H$\alpha$ and X-ray emissions of the wind simulated for a 0.01~kpc off-centre continuous starburst (MOC10c)  at t=4~Myr}
    \label{fig:OC10SSCHalpha}
\end{figure}

We also performed a third simulation with model MOC10c, in which the starburst ring is shifted slightly to the north, by only 10~pc above the galactic plane. In this case (see Fig. \ref{fig:OC10SSCHalpha}), the asymmetry between the northern and southern sides is less pronounced than in model MOC100c. The main difference lies in the forward shock propagating into the galactic halo: on the southern side, it appears slightly flatter and extends only to about 5~kpc, compared to 6~kpc on the northern side.

\begin{figure}[!h]
    \centering
    \includegraphics[width=\columnwidth]{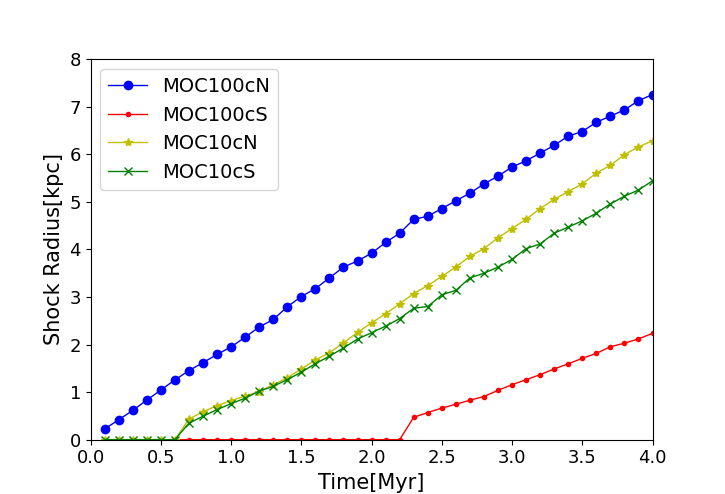}
    \caption[]{Radius of the halo forward shock generated by the flow in both the north and south sides of the wind in the MOC10C and MOC100c models against time, considering the distance from the centre of the galaxy to the highest point of the shock.}
    \label{fig:shockradi}
\end{figure}

Figure~\ref{fig:shockradi} shows the radial position of the halo forward shock on the northern and southern sides for models MOC10c and MOC100c, considering the distance from the centre of the galaxy to the highest point of the shock and measuring only from the moment that it breaks out of the disk. In model MOC10c, up to t=$0.6$~Myr, both the northern and southern bubbles grow symmetrically within the galactic disk. They begin to break out around t=$1.4$~Myr, with only a slight difference in timing. During this breakout phase, the amount of galactic gas swept up by the wind in the southern-side is slightly greater than on the northern side. Additionally, the bubble extends laterally by about 0.1~kpc in the south, which further increases the swept-up mass in that direction. This enhanced mass loading slows down the propagation of the forward shock on the southern side. This result demonstrates that even a small vertical displacement of the starburst region can produce a significative asymmetry in the resulting galactic wind. This asymmetry might not be immediately apparent when considering only its size at a single point in time, but it becomes evident  along the time evolution.

\begin{figure}[!h]
    \centering
   \includegraphics[width=\columnwidth]{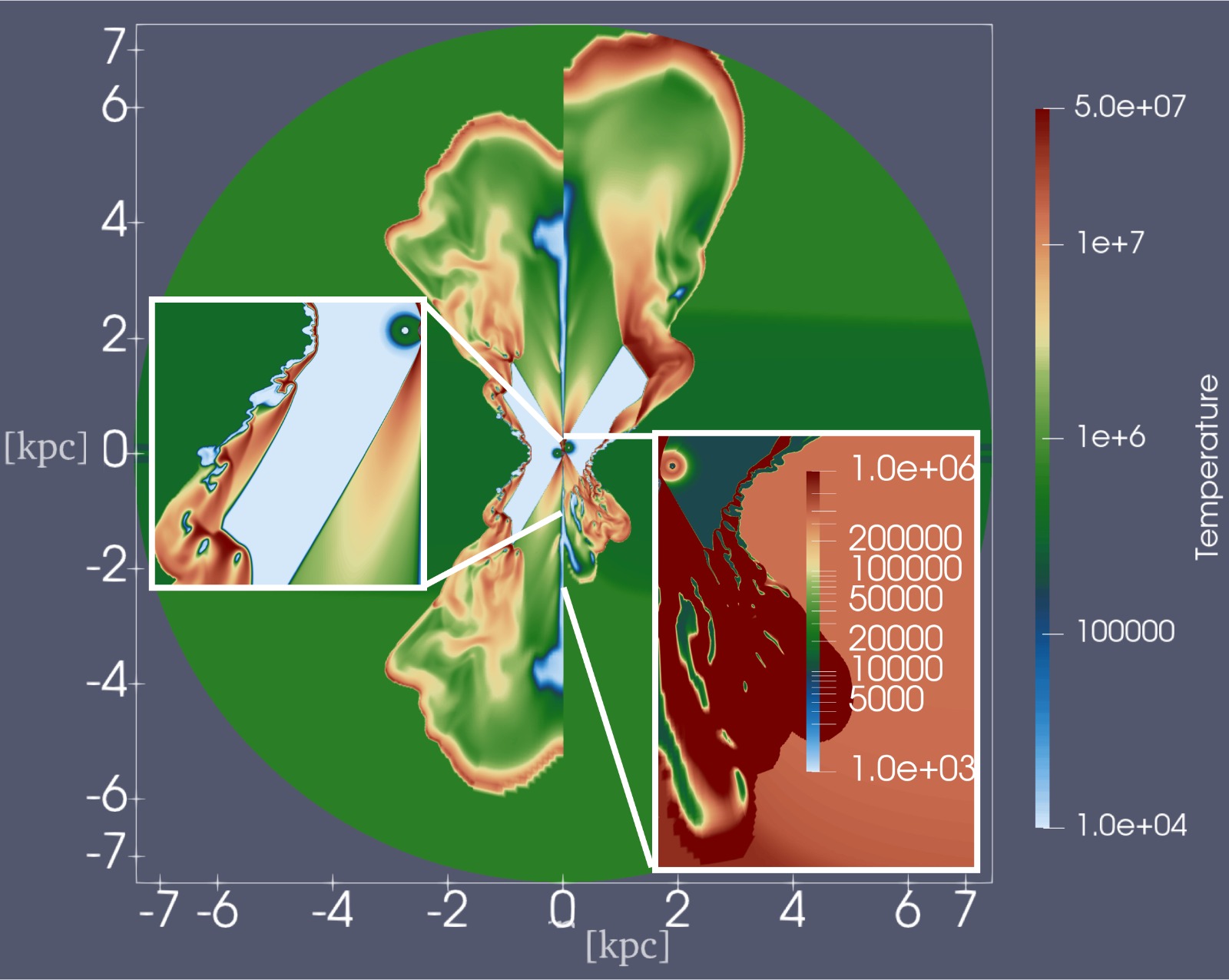}
    \caption[]{Temperature map for the models MCENc (left side) and the MOC100c (right side), both at a time of 4~Myr.}
    \label{fig:Temperature}
\end{figure}

The multi-phase structure of the galactic wind bubble, both for the model with the starburst cluster in the equatorial plane and for the model with the starburst cluster shifted to the north, is illustrated in the temperature maps shown in Fig.~\ref{fig:Temperature}. The free wind region, with a temperature of the order of $10^4$~K, extends conically outwards from the starburst ring. It surrounds the inner shocked wind, which converges towards the galactic axis and is deflected into a warmer flow with temperatures around $10^7$~K. This temperature gradually decreases with height as the thermal energy is converted into kinetic energy.

At the outer edge of the free-wind region, the wind deflected by the galactic disk forms a dense, hot, thin shell with temperatures also around $10^7$~K. This shell maintains its high temperature up to a height of $\sim 2$~kpc. This is due to the lateral compression by the dense galactic disk and the internal pressure of the galactic wind. Despite strong radiative losses, the continuous injection of momentum from the galactic wind maintains the high temperature of the shell. This dense shell is responsible for significant X-ray emission and has filamentary structures. In the model with the displaced starburst ring, this shell extends further upwards on the northern side and appears more filamentary and fragmented on the southern side.

\begin{figure}[!h]
    \centering
    \includegraphics[width=\columnwidth]{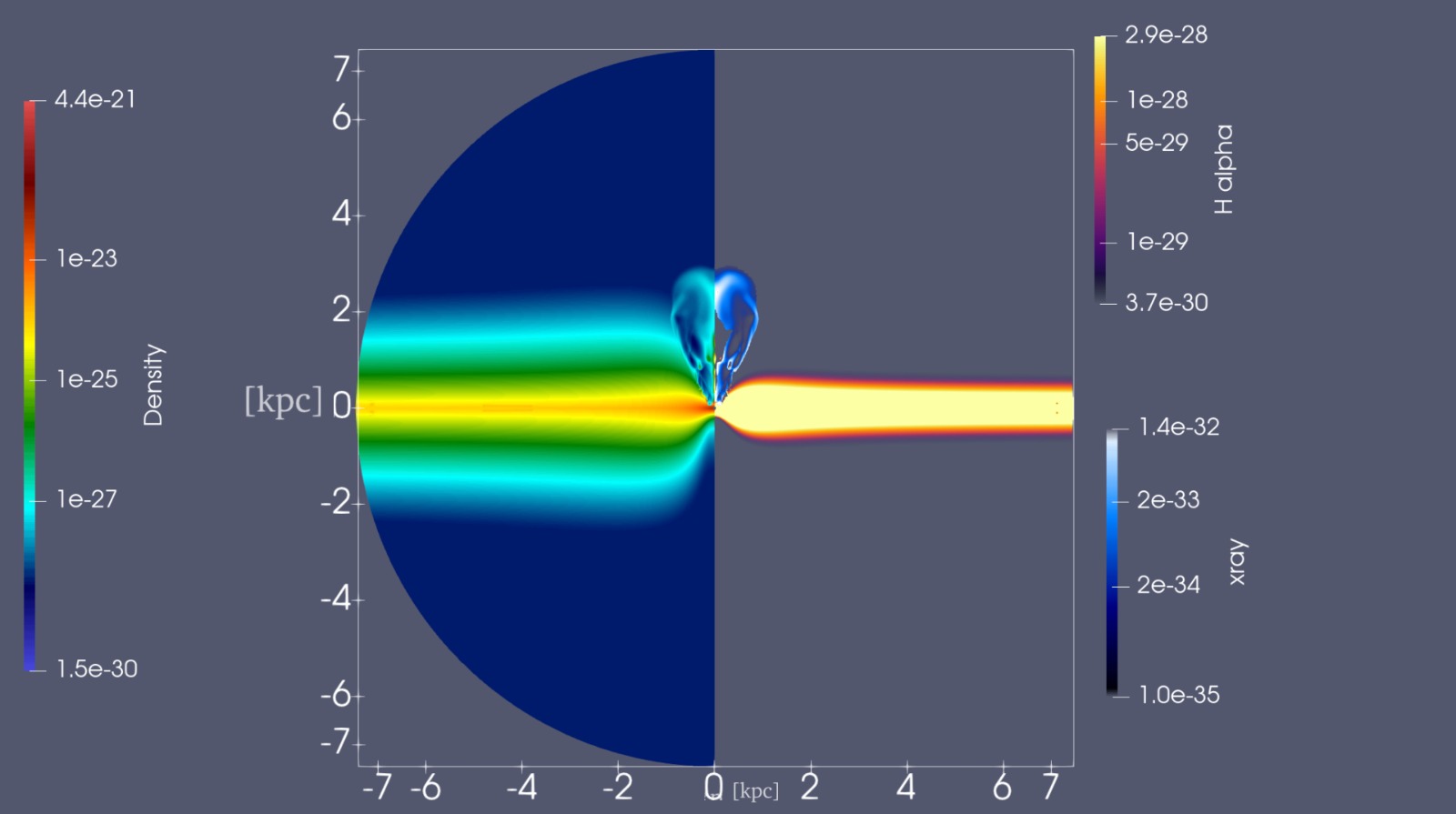}
    \caption[]{H$\alpha$ and X-ray emissions of the wind simulated for a 0.1~kpc off-centre instantaneous starburst (MOC100i)  at t=4~Myr}
    \label{fig:OC2InsHalpha}
\end{figure}

In all the models discussed above, we have considered continuous starburst injection. However, we also performed a simulation with an instantaneous starburst, offset by 0.1~kpc from the galactic centre (model MOC100i). As shown in Fig.\ref{fig:OC2InsHalpha}, this model produces a galactic bubble that develops only on the northern-side of the galaxy, this behavior persists even at the maximum time of the simulation (10~Myr). Moreover, even in the northern side, the galactic bubble remains small, 
with high-density regions appearing only in the gas swept up by the main shock. This is because even though we are injecting more energy during the first 1.3~Myr of the simulation, this  leads to an increase in the ram pressure, causing the bubble to evolve more slowly than in the continuous models.

\begin{figure}[!h]
    \centering
    \includegraphics[width=\columnwidth]{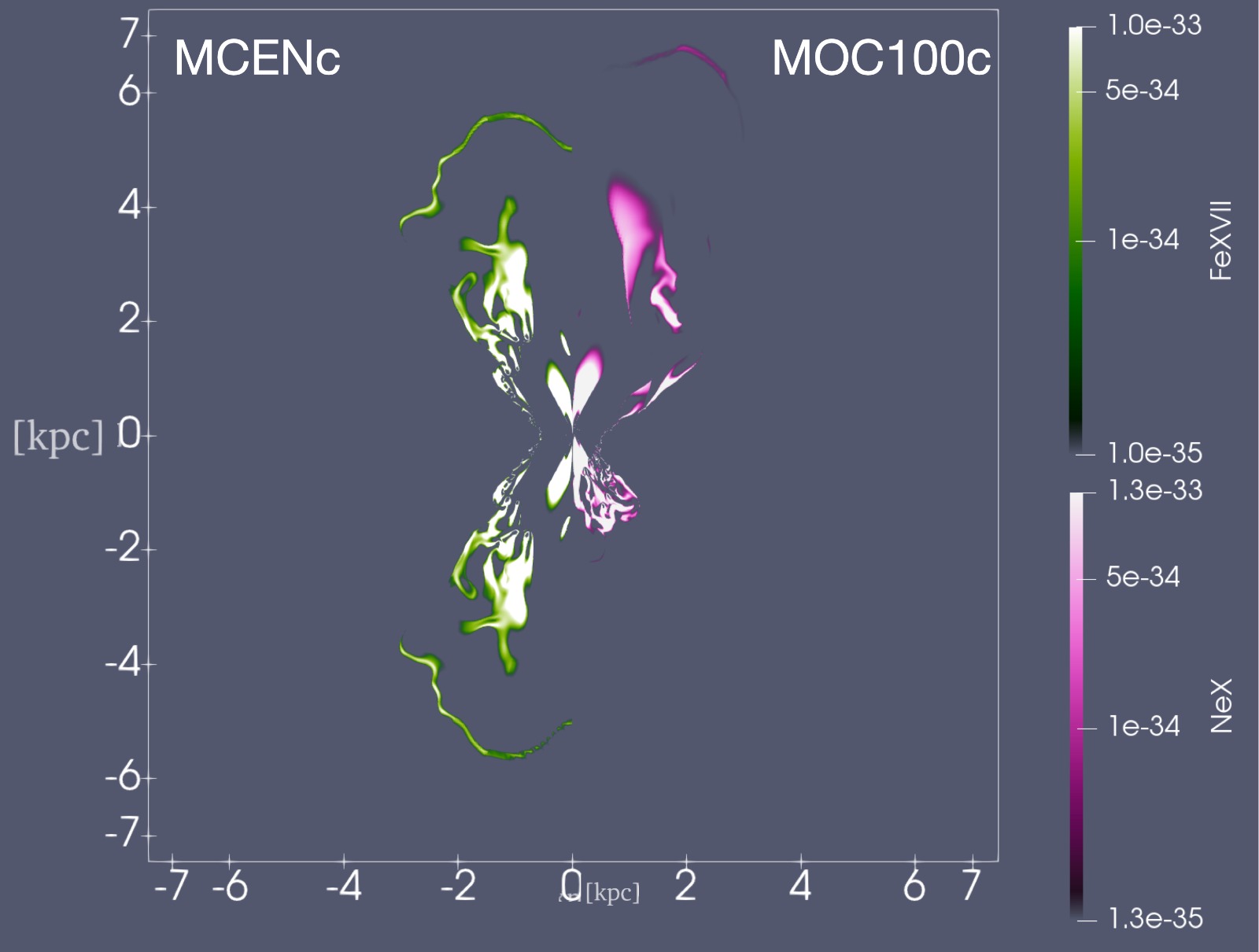}
    \caption[]{X-ray emissions for the Fe~XVII   line in the MCENc model (left side) and the Ne~X line for the MOC100c model (right side), both at a time of 4~Myr.}
    \label{fig:OC2CENFeN}
\end{figure}

In Fig. \ref{fig:OC2CENFeN}, we present maps of the emission for the two most prominent lines present in the range of temperatures of the X-ray previously presented  emission
maps: Fe~XVII 1.5~nm and Ne~X 1.02~nm. These lines were obtained using the emission coefficients of \cite{SandersFabian2011MNRAS} and the abundances for a solar metallicity of \cite{Grevesse1989AIPC}. We can appreciate a more prominent emission of Ne~X in the free wind zone, while  Fe~XVII is mainly better at tracing the halo forward shocks.

\subsection{Mass and metallicity loss}

\begin{figure}[!h]
    \centering
    \includegraphics[width=\columnwidth]{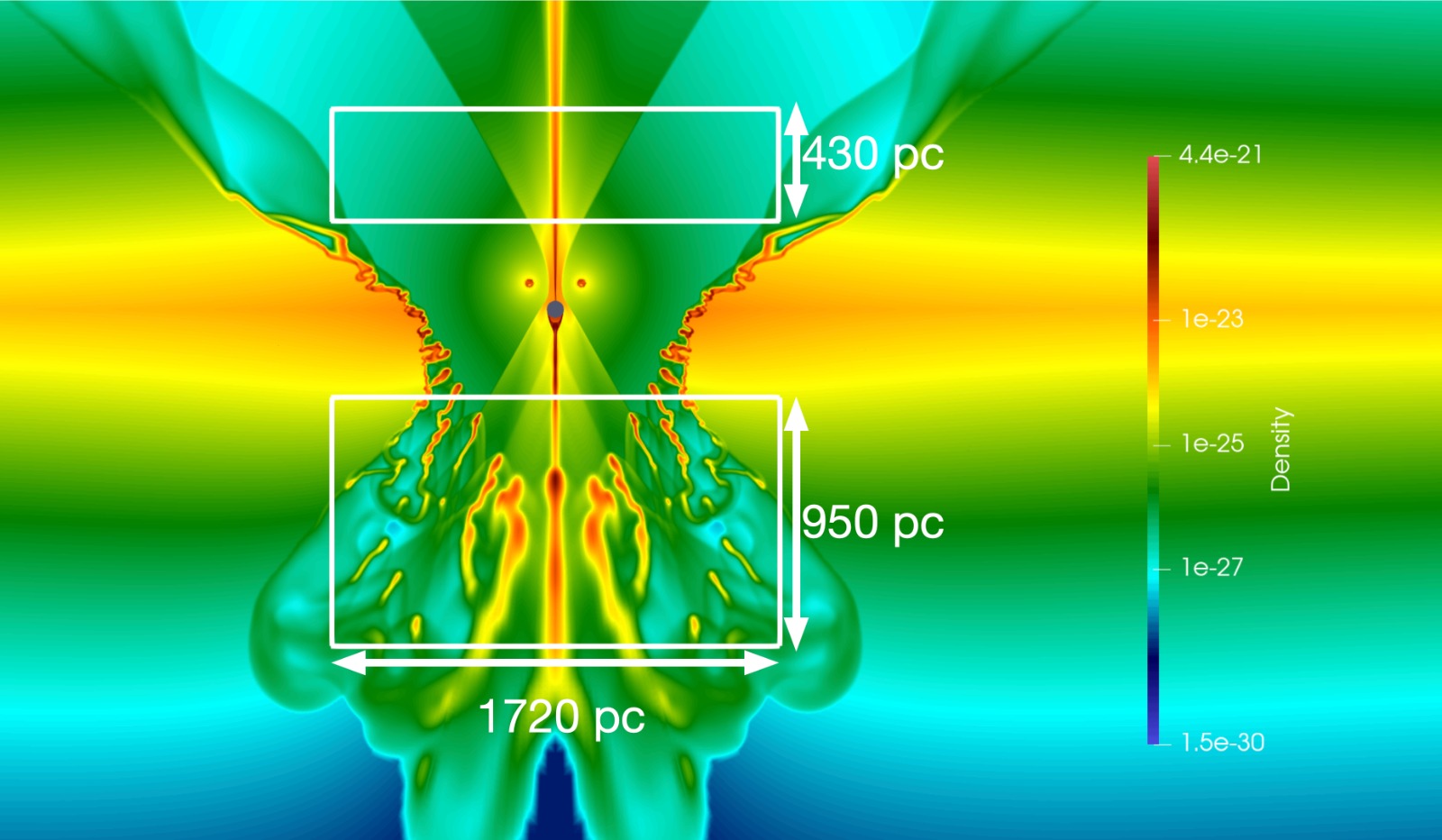}
    \caption{Mass analysis grid used to measure mass flux, analogous to the one used in \cite{Lopez2023}.}
    \label{fig:grid}
\end{figure}

To compare the mass fluxes of our models with those reported in \cite{Lopez2023}, we computed the mass flux in the northern and southern lobes of each model, 
using the same areas of analysis used in that paper (Fig. \ref{fig:grid}). For the northern hemisphere, we took an area of 1.72~kpc by 430~pc at a height of 340~pc from the centre of the galaxy's disk. For the southern hemisphere, we took an area of 1.72~kpc by 950~pc at a height of -340~pc. 

\begin{figure}[!h]
    \centering
    \includegraphics[width=\columnwidth]{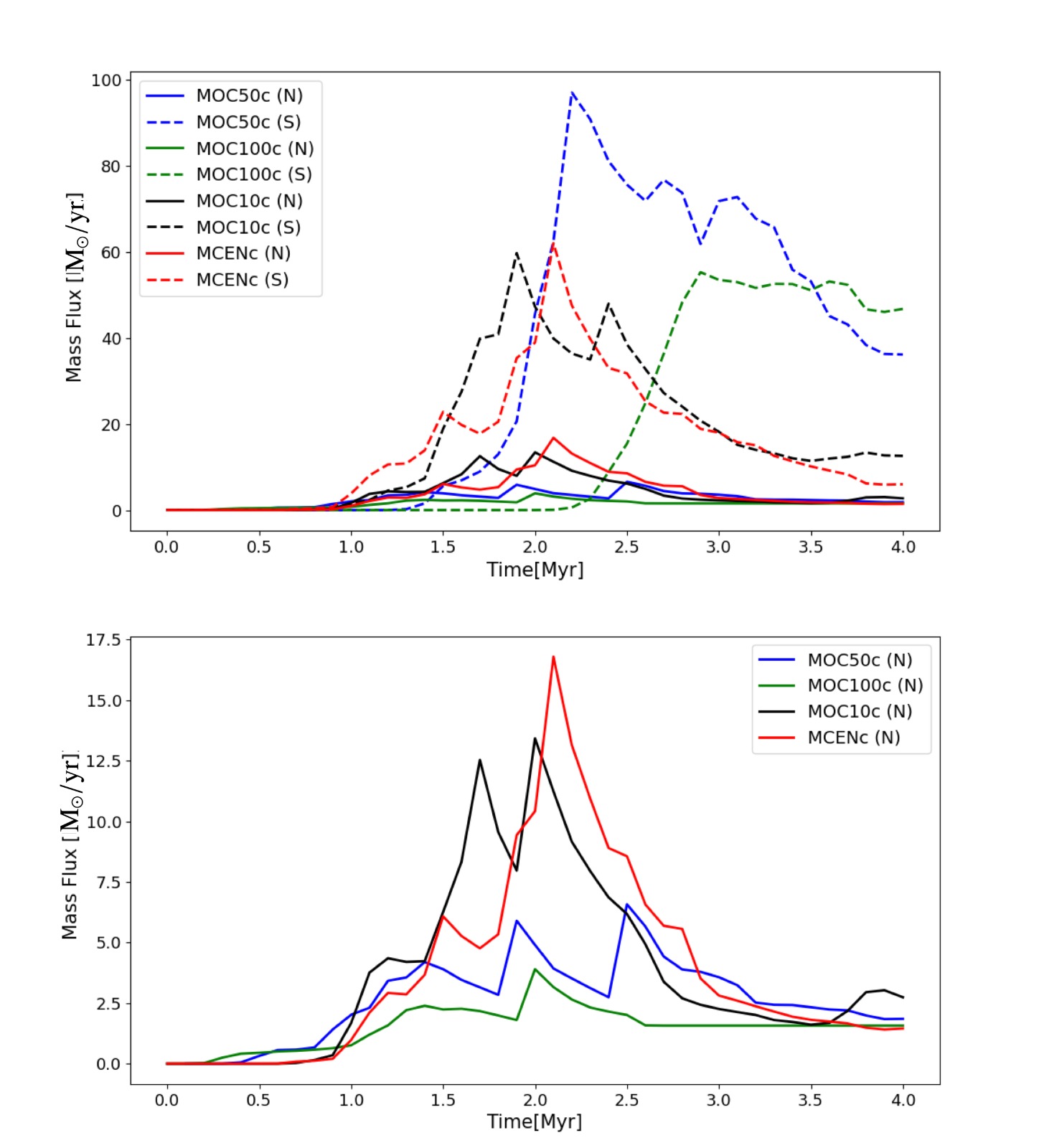}
    \caption[]{Upper image: Mass flux against time for the northern (continuous line) and southern (doted line) regions for the models with continuous starburst injection. Lower image: Zoom on the upper image,  presenting the northern regions only. }
    \label{fig:MvT}
\end{figure}

Figure~\ref{fig:MvT} displays the time evolution of the mass flux, showing several local maxima and significant differences between the northern and southern components.

Starting with the model where the  starburst ring is located at the galactic plane, we observed a symmetrical behavior between the two lobes, differing primarily in magnitude. This is expected, as the wind flux is intrinsically similar, and the variation arises due to differences in the analysis area. The mass flux reaches a peak around 2~Myr, with values of approximately 60~M$_{\odot}$/yr in the southern lobe and 17~M$_{\odot}$/yr in the northern one. After this peak, the dragged mass decreases, with the northern flux stabilising around 2~M$_{\odot}$/yr. Multiple local maxima are evident, likely caused by the passage of dense structures formed within the wind.

In the off-plane models, the northern lobe behaves more like a free wind, reaching a maximum mass flux of about 13.5~M$_\odot$/yr. For example, in the 10~pc offset model, this peak occurs at approximately 2~Myr. Afterward, the mass flux declines toward a stable value. In contrast, the southern lobes show no mass flux at the beginning of the simulation. The flux begins around 1~Myr for the 10~pc model, 1.3~Myr for the 50~pc model, and 2.1~Myr for the 100~pc model. Following the onset, the southern mass flux increases rapidly, reaches a global maximum, and then declines, also tending toward a free-wind regime.

It is important to note that although the global maximal can reach values as high as 97~M$_\odot$/yr, these high mass-loss rates are short-lived, lasting on a timescale of the order of 0.5~Myr. This short time window is small relative to the overall lifetime of a starburst, which may limit the observational likelihood of detecting such extreme outflows, depending on the evolutionary phase.

These results have a difference of up to an order of magnitude with respect to the measured values of \cite{Lopez2023}, 2.8~M$_{\odot}$~yr$^{-1}$ for the northern outflow and 3.2~M$_{\odot}$~yr$^{-1}$ for the southern one. This is due in part to the knots of the central density. However, the majority of the mass causing the disparity of the south flow comes from the filaments; also, it is important to note that given the way that we 
performed the integration (i.e. just as a rotation of the axisymmetric simulation), each of the filaments was considered as a continuous surface instead of the multiple thin filaments that we would generate with a 3D simulation, which could raise considerably the mass flux. However, we should also consider that the difference of an order of magnitude only lasts for a short period of time. We can get a better match with the observations if we consider that the wind on NGC~253 would be outside the maximum ejection peak; for instance, as an earlier south side ejection or a late type where the northern flow is already in a free wind state and the southern one has passed its maximum (but  not yet reaching the status of a stable free-wind). For our models, we find that the best fit is the model  MOC100c at 2.3~Myr, the mass flux values vary by 0.5~M$_{\odot}/$yr from those reported as the closest to the observations.

\section{Conclusions}

We carried out hydrodynamic simulations for winds generated by starburst, considering a RSS and a host galaxy conforming to a disk and a halo structure. Our numerical models considered the evolution of the starburst first as an instantaneous and then as a continuous one. We also took into account the altitude position of the starburst ring. We did not consider galaxy rotation, dust dynamics, the effect of gravity, or magnetic fields. Our results show that
the galactic wind from a ring with continuous star formation presents a more complex structure than previously described models considering a spherical burst. This structure, particularly the well-defined cone of interjection on the wind, break through the disk, generating a north and south side, even when starting of-centre. The interaction of the winds and the disk of the galaxies produce filamentary and dense structures with H$\alpha$ emission, with lengths exceeding one or two kpc, comparable to observed structures of around 2~kpc \cite{ShopbelandBland1998}.

The non-nuclear star formation burst off-centre creates a significant difference in length, morphology, and mass flow between the northern and southern outflows. The interaction of the galactic wind with the dense filamentary structures induces internal shocks within the wind region, resulting in thermal X-ray emissions distributed throughout the wind, rather than remaining confined to the shocks. 

In the context of a 'thick disk' and a vertical displacement of less than 100 pc, our models indicate that the mass flow between the northern and southern outflows can reach a difference of an order of magnitude. 
When we consider the evolution of the mass flux over time, there are windows of time around these maximal points where the observed mass flux \citep{Lopez2023} can be replicated; for our models, we found the best result when considering a 100~pc distance from the centre and an age of 2.3~Myr. However, further studies on the mass dragged in the central overdensity are needed. 

We plan to perform complementary tridimensional simulations that will help us to clarify the behaviour of the filaments in the wind, as well as the presence of the central overdensity, enabling us to perform a more refined study of the mass flux .

\begin{acknowledgements} 

We acknowledge support of the UNAM-PAPIIT grants IN110722 and AG101125. J.A.O.-C. acknowledges scholarship from SECIHTI-México.  A.R.G. is grateful to DGAPA-UNAM for the grants during an academic stay in LUTH, Observatoire de Paris, and is also grateful to LUTH, Observatoire de Paris for their hospitality.

\end{acknowledgements}

\bibliographystyle{aa.bst} 
\bibliography{biblio.bib}

@ARTICLE{Chen_Oh_2024MNRAS.530.4032C,
       author = {{Chen}, Zirui and {Oh}, S. Peng},
        title = "{The survival and entrainment of molecules and dust in galactic winds}",
      journal = {\mnras},
         year = 2024,
        month = jun,
       volume = {530},
       number = {4},
        pages = {4032-4057},
          doi = {10.1093/mnras/stae1113},
archivePrefix = {arXiv},
       eprint = {2311.04275},
 primaryClass = {astro-ph.GA},
       adsurl = {https://ui.adsabs.harvard.edu/abs/2024MNRAS.530.4032C}
}

@ARTICLE{Chevalier1985,
       author = {{Chevalier}, R.~A. and {Clegg}, A.~W.},
        title = "{Wind from a starburst galaxy nucleus}",
      journal = {\nat},
         year = 1985,
        month = sep,
       volume = {317},
       number = {6032},
        pages = {44-45},
          doi = {10.1038/317044a0},
       adsurl = {https://ui.adsabs.harvard.edu/abs/1985Natur.317...44C}
}

@ARTICLE{Garnett_2002ApJ...581.1019G,
       author = {{Garnett}, Donald R.},
        title = "{The Luminosity-Metallicity Relation, Effective Yields, and Metal Loss in Spiral and Irregular Galaxies}",
      journal = {\apj},
         year = 2002,
        month = dec,
       volume = {581},
       number = {2},
        pages = {1019-1031},
          doi = {10.1086/344301},
archivePrefix = {arXiv},
       eprint = {astro-ph/0209012},
 primaryClass = {astro-ph},
       adsurl = {https://ui.adsabs.harvard.edu/abs/2002ApJ...581.1019G}
}

@ARTICLE{Hirschauer_etal_2018AJ....155...82H,
       author = {{Hirschauer}, Alec S. and {Salzer}, John J. and {Janowiecki}, Steven and {Wegner}, Gary A.},
        title = "{Metal Abundances of KISS Galaxies. VI. New Metallicity Relations for the KISS Sample of Star-forming Galaxies}",
      journal = {\aj},
         year = 2018,
        month = feb,
       volume = {155},
       number = {2},
          eid = {82},
        pages = {82},
          doi = {10.3847/1538-3881/aaa4ba},
archivePrefix = {arXiv},
       eprint = {1801.01133},
 primaryClass = {astro-ph.GA},
       adsurl = {https://ui.adsabs.harvard.edu/abs/2018AJ....155...82H}
}

@INCOLLECTION{Heckman_Thompson_2017hsn..book.2431H,
       author = {{Heckman}, Timothy M. and {Thompson}, Todd A.},
        title = "{Galactic Winds and the Role Played by Massive Stars}",
    booktitle = {Handbook of Supernovae},
         year = 2017,
       editor = {{Alsabti}, Athem W. and {Murdin}, Paul},
        pages = {2431},
          doi = {10.1007/978-3-319-21846-5_23},
       adsurl = {https://ui.adsabs.harvard.edu/abs/2017hsn..book.2431H}
}

@ARTICLE{Knapen_2005A&A...429..141K,
       author = {{Knapen}, J.~H.},
        title = "{Structure and star formation in disk galaxies. III. Nuclear and circumnuclear H{\ensuremath{\alpha}} emission}",
      journal = {\aap},
         year = 2005,
        month = jan,
       volume = {429},
        pages = {141-151},
          doi = {10.1051/0004-6361:20041909},
archivePrefix = {arXiv},
       eprint = {astro-ph/0409031},
 primaryClass = {astro-ph},
       adsurl = {https://ui.adsabs.harvard.edu/abs/2005A&A...429..141K}
}

@ARTICLE{Keppens2021,
       author = {{Keppens}, Rony and {Teunissen}, Jannis and {Xia}, Chun and {Porth}, Oliver},
        title = "{MPI-AMRVAC: a parallel, grid-adaptive PDE toolkit}",
      journal = {arXiv e-prints},
         year = 2020,
        month = apr,
          eid = {arXiv:2004.03275},
        pages = {arXiv:2004.03275},
archivePrefix = {arXiv},
       eprint = {2004.03275},
 primaryClass = {astro-ph.IM},
       adsurl = {https://ui.adsabs.harvard.edu/abs/2020arXiv200403275K}
}

@ARTICLE{Li_etal_2015ApJS..216....6L,
       author = {{Li}, Shuo and {de Grijs}, Richard and {Anders}, Peter and {Li}, Chengyuan},
        title = "{Star Cluster Disruption in the Starburst Galaxy Messier 82}",
      journal = {\apjs},
         year = 2015,
        month = jan,
       volume = {216},
       number = {1},
          eid = {6},
        pages = {6},
          doi = {10.1088/0067-0049/216/1/6},
archivePrefix = {arXiv},
       eprint = {1411.2704},
 primaryClass = {astro-ph.GA},
       adsurl = {https://ui.adsabs.harvard.edu/abs/2015ApJS..216....6L}
}

@ARTICLE{Levy_etal_2024ApJ...973L..55L,
       author = {{Levy}, Rebecca C. and {Bolatto}, Alberto D. and {Mayya}, Divakara and {Cuevas-Otahola}, Bolivia and {Tarantino}, Elizabeth and {Boyer}, Martha L. and {Boogaard}, Leindert A. and {B{\"o}ker}, Torsten and {Cronin}, Serena A. and {Dale}, Daniel A. and {Donaghue}, Keaton and {Emig}, Kimberly L. and {Fisher}, Deanne B. and {Glover}, Simon C.~O. and {Herrera-Camus}, Rodrigo and {Jim{\'e}nez-Donaire}, Mar{\'\i}a J. and {Klessen}, Ralf S. and {Lenki{\'c}}, Laura and {Leroy}, Adam K. and {De Looze}, Ilse and {Meier}, David S. and {Mills}, Elisabeth A.~C. and {Ott}, Juergen and {Rela{\~n}o}, M{\'o}nica and {Veilleux}, Sylvain and {Villanueva}, Vicente and {Walter}, Fabian and {van der Werf}, Paul P.},
        title = "{JWST Observations of Starbursts: Massive Star Clusters in the Central Starburst of M82}",
      journal = {\apjl},
         year = 2024,
        month = oct,
       volume = {973},
       number = {2},
          eid = {L55},
        pages = {L55},
          doi = {10.3847/2041-8213/ad7af3},
archivePrefix = {arXiv},
       eprint = {2408.04135},
 primaryClass = {astro-ph.GA},
       adsurl = {https://ui.adsabs.harvard.edu/abs/2024ApJ...973L..55L}
}

@ARTICLE{Mayya_etal_2008ApJ...679..404M,
       author = {{Mayya}, Y.~D. and {Romano}, R. and {Rodr{\'\i}guez-Merino}, L.~H. and {Luna}, A. and {Carrasco}, L. and {Rosa-Gonz{\'a}lez}, D.},
        title = "{HST ACS Imaging of M82: A Comparison of Mass and Size Distribution Functions of the Younger Nuclear and Older Disk Clusters}",
      journal = {\apj},
         year = 2008,
        month = may,
       volume = {679},
       number = {1},
        pages = {404-419},
          doi = {10.1086/587541},
archivePrefix = {arXiv},
       eprint = {0802.1922},
 primaryClass = {astro-ph},
       adsurl = {https://ui.adsabs.harvard.edu/abs/2008ApJ...679..404M}
}

@ARTICLE{Mazzuca_etal_2008ApJS..174..337M,
       author = {{Mazzuca}, Lisa M. and {Knapen}, Johan H. and {Veilleux}, Sylvain and {Regan}, Michael W.},
        title = "{A Connection between Star Formation in Nuclear Rings and Their Host Galaxies}",
      journal = {\apjs},
         year = 2008,
        month = feb,
       volume = {174},
       number = {2},
        pages = {337-365},
          doi = {10.1086/522338},
archivePrefix = {arXiv},
       eprint = {0709.2915},
 primaryClass = {astro-ph},
       adsurl = {https://ui.adsabs.harvard.edu/abs/2008ApJS..174..337M}
}

@ARTICLE{Miyamoto_Nagai1975PASJ...27..533M,
       author = {{Miyamoto}, M. and {Nagai}, R.},
        title = "{Three-dimensional models for the distribution of mass in galaxies.}",
      journal = {\pasj},
         year = 1975,
        month = jan,
       volume = {27},
        pages = {533-543},
       adsurl = {https://ui.adsabs.harvard.edu/abs/1975PASJ...27..533M}
}

@ARTICLE{Portegies_Zwart_etal2010ARA&A..48..431P,
       author = {{Portegies Zwart}, Simon F. and {McMillan}, Stephen L.~W. and {Gieles}, Mark},
        title = "{Young Massive Star Clusters}",
      journal = {\araa},
         year = 2010,
        month = sep,
       volume = {48},
        pages = {431-493},
          doi = {10.1146/annurev-astro-081309-130834},
archivePrefix = {arXiv},
       eprint = {1002.1961},
 primaryClass = {astro-ph.GA},
       adsurl = {https://ui.adsabs.harvard.edu/abs/2010ARA&A..48..431P}
}

@ARTICLE{Rubin_etal_2014ApJ...794..156R,
       author = {{Rubin}, Kate H.~R. and {Prochaska}, J. Xavier and {Koo}, David C. and {Phillips}, Andrew C. and {Martin}, Crystal L. and {Winstrom}, Lucas O.},
        title = "{Evidence for Ubiquitous Collimated Galactic-scale Outflows along the Star-forming Sequence at z \raisebox{-0.5ex}\textasciitilde 0.5}",
      journal = {\apj},
         year = 2014,
        month = oct,
       volume = {794},
       number = {2},
          eid = {156},
        pages = {156},
          doi = {10.1088/0004-637X/794/2/156},
archivePrefix = {arXiv},
       eprint = {1307.1476},
 primaryClass = {astro-ph.CO},
       adsurl = {https://ui.adsabs.harvard.edu/abs/2014ApJ...794..156R}
}

@ARTICLE{Ryon_etal_2017ApJ...841...92R,
       author = {{Ryon}, J.~E. and {Gallagher}, J.~S. and {Smith}, L.~J. and {Adamo}, A. and {Calzetti}, D. and {Bright}, S.~N. and {Cignoni}, M. and {Cook}, D.~O. and {Dale}, D.~A. and {Elmegreen}, B.~E. and {Fumagalli}, M. and {Gouliermis}, D.~A. and {Grasha}, K. and {Grebel}, E.~K. and {Kim}, H. and {Messa}, M. and {Thilker}, D. and {Ubeda}, L.},
        title = "{Effective Radii of Young, Massive Star Clusters in Two LEGUS Galaxies}",
      journal = {\apj},
         year = 2017,
        month = jun,
       volume = {841},
       number = {2},
          eid = {92},
        pages = {92},
          doi = {10.3847/1538-4357/aa719e},
archivePrefix = {arXiv},
       eprint = {1705.02692},
 primaryClass = {astro-ph.GA},
       adsurl = {https://ui.adsabs.harvard.edu/abs/2017ApJ...841...92R}
}

@ARTICLE{ary2008,
       author = {{Rodr{\'\i}guez-Gonz{\'a}lez}, A. and {Esquivel}, A. and {Vel{\'a}zquez}, P.~F. and {Raga}, A.~C. and {Melo}, V.},
        title = "{Filaments in Galactic Winds Driven by Young Stellar Clusters}",
      journal = {\apj},
         year = 2008,
        month = dec,
       volume = {689},
       number = {1},
        pages = {153-159},
          doi = {10.1086/592492},
archivePrefix = {arXiv},
       eprint = {0808.1611},
 primaryClass = {astro-ph},
       adsurl = {https://ui.adsabs.harvard.edu/abs/2008ApJ...689..153R}
}

@ARTICLE{Thompson_Heckman_2024ARA&A..62..529T,
       author = {{Thompson}, Todd A. and {Heckman}, Timothy M.},
        title = "{Theory and Observation of Winds from Star-Forming Galaxies}",
      journal = {\araa},
         year = 2024,
        month = sep,
       volume = {62},
       number = {1},
        pages = {529-591},
          doi = {10.1146/annurev-astro-041224-011924},
archivePrefix = {arXiv},
       eprint = {2406.08561},
 primaryClass = {astro-ph.GA},
       adsurl = {https://ui.adsabs.harvard.edu/abs/2024ARA&A..62..529T}
}

@ARTICLE{Sanders_etal_2023ApJ...942...24S,
       author = {{Sanders}, Ryan L. and {Shapley}, Alice E. and {Jones}, Tucker and {Shivaei}, Irene and {Popping}, Gerg{\"o} and {Reddy}, Naveen A. and {Dav{\'e}}, Romeel and {Price}, Sedona H. and {Mobasher}, Bahram and {Kriek}, Mariska and {Coil}, Alison L. and {Siana}, Brian},
        title = "{CO Emission, Molecular Gas, and Metallicity in Main-sequence Star-forming Galaxies at z {\ensuremath{\sim}} 2.3}",
      journal = {\apj},
         year = 2023,
        month = jan,
       volume = {942},
       number = {1},
          eid = {24},
        pages = {24},
          doi = {10.3847/1538-4357/aca46f},
archivePrefix = {arXiv},
       eprint = {2204.06937},
 primaryClass = {astro-ph.GA},
       adsurl = {https://ui.adsabs.harvard.edu/abs/2023ApJ...942...24S}
}

@ARTICLE{Strickland-Stevens-2000,
       author = {{Strickland}, David K. and {Stevens}, Ian R.},
        title = "{Starburst-driven galactic winds - I. Energetics and intrinsic X-ray emission}",
      journal = {\mnras},
         year = 2000,
        month = may,
       volume = {314},
       number = {3},
        pages = {511-545},
          doi = {10.1046/j.1365-8711.2000.03391.x},
archivePrefix = {arXiv},
       eprint = {astro-ph/0001395},
 primaryClass = {astro-ph},
       adsurl = {https://ui.adsabs.harvard.edu/abs/2000MNRAS.314..511S}
}

@ARTICLE{Strickland_2002,
       author = {{Strickland}, David K. and {Heckman}, Timothy M. and {Weaver}, Kimberly A. and {Hoopes}, Charles G. and {Dahlem}, Michael},
        title = "{Chandra Observations of NGC 253. II. On the Origin of Diffuse X-Ray Emission in the Halos of Starburst Galaxies}",
      journal = {\apj},
         year = 2002,
        month = apr,
       volume = {568},
       number = {2},
        pages = {689-716},
          doi = {10.1086/338889},
archivePrefix = {arXiv},
       eprint = {astro-ph/0111511},
 primaryClass = {astro-ph},
       adsurl = {https://ui.adsabs.harvard.edu/abs/2002ApJ...568..689S}
}

@ARTICLE{Schure_etal_2009A&A...508..751S,
       author = {{Schure}, K.~M. and {Kosenko}, D. and {Kaastra}, J.~S. and {Keppens}, R. and {Vink}, J.},
        title = "{A new radiative cooling curve based on an up-to-date plasma emission code}",
      journal = {\aap},
         year = 2009,
        month = dec,
       volume = {508},
       number = {2},
        pages = {751-757},
          doi = {10.1051/0004-6361/200912495},
archivePrefix = {arXiv},
       eprint = {0909.5204},
 primaryClass = {astro-ph.GA},
       adsurl = {https://ui.adsabs.harvard.edu/abs/2009A&A...508..751S}
}

@ARTICLE{VanMarleKeppens_2011CF.....42...44V,
       author = {{van Marle}, Allard Jan and {Keppens}, Rony},
        title = "{Radiative cooling in numerical astrophysics: The need for adaptive mesh refinement}",
      journal = {Computers and Fluids},
         year = 2011,
        month = mar,
       volume = {42},
       number = {1},
        pages = {44-53},
          doi = {10.1016/j.compfluid.2010.10.022},
archivePrefix = {arXiv},
       eprint = {1011.2610},
 primaryClass = {astro-ph.IM},
       adsurl = {https://ui.adsabs.harvard.edu/abs/2011CF.....42...44V}
}

@ARTICLE{Weedman_etal_1981ApJ...248..105W,
       author = {{Weedman}, D.~W. and {Feldman}, F.~R. and {Balzano}, V.~A. and {Ramsey}, L.~W. and {Sramek}, R.~A. and {Wuu}, C. -C.},
        title = "{NGC 7714 - The prototype star-burst galactic nucleus}",
      journal = {\apj},
         year = 1981,
        month = aug,
       volume = {248},
        pages = {105-112},
          doi = {10.1086/159133},
       adsurl = {https://ui.adsabs.harvard.edu/abs/1981ApJ...248..105W}
}

@ARTICLE{zhang2018,
       author = {{Zhang}, Dong},
        title = "{A Review of the Theory of Galactic Winds Driven by Stellar Feedback}",
      journal = {Galaxies},
         year = 2018,
        month = nov,
       volume = {6},
       number = {4},
        pages = {114},
          doi = {10.3390/galaxies6040114},
archivePrefix = {arXiv},
       eprint = {1811.00558},
 primaryClass = {astro-ph.GA},
       adsurl = {https://ui.adsabs.harvard.edu/abs/2018Galax...6..114Z}
}

@ARTICLE{Holtzman1992,
       author = {{Holtzman}, J.~A. and {Faber}, S.~M. and {Shaya}, E.~J. and {Lauer}, T.~R. and {Groth}, J. and {Hunter}, D.~A. and {Baum}, W.~A. and {Ewald}, S.~P. and {Hester}, J.~J. and {Light}, R.~M. and {Lynds}, C.~R. and {O'Neil}, E.~J., Jr. and {Westphal}, J.~A.},
        title = "{Planetary Camera Observations of NGC 1275: Discovery of a Central Population of Compact Massive Blue Star Clusters}",
      journal = {\aj},
         year = 1992,
        month = mar,
       volume = {103},
        pages = {691},
          doi = {10.1086/116094},
       adsurl = {https://ui.adsabs.harvard.edu/abs/1992AJ....103..691H}
}

@ARTICLE{Melo2005,
       author = {{Melo}, V.~P. and {Mu{\~n}oz-Tu{\~n}{\'o}n}, C. and {Ma{\'i}z-Apell{\'a}niz}, J. and {Tenorio-Tagle}, G.},
        title = "{Erratum: ``Young Super Star Clusters in the Starburst of M82: The Catalogue`` <A href=''/abs/2005ApJ...619..270M''>ApJ, 619, 270 [2005]</A>)}",
      journal = {\apj},
         year = 2005,
        month = oct,
       volume = {632},
       number = {1},
        pages = {684-688},
          doi = {10.1086/432751},
       adsurl = {https://ui.adsabs.harvard.edu/abs/2005ApJ...632..684M}
}

@ARTICLE{DErcole1999,
       author = {{D'Ercole}, A. and {Brighenti}, F.},
        title = "{Galactic winds and circulation of the interstellar medium in dwarf galaxies}",
      journal = {\mnras},
         year = 1999,
        month = nov,
       volume = {309},
       number = {4},
        pages = {941-954},
          doi = {10.1046/j.1365-8711.1999.02911.x},
archivePrefix = {arXiv},
       eprint = {astro-ph/9907005},
 primaryClass = {astro-ph},
       adsurl = {https://ui.adsabs.harvard.edu/abs/1999MNRAS.309..941D}
}

@ARTICLE{MacLow1999,
       author = {{Mac Low}, Mordecai-Mark and {Ferrara}, Andrea},
        title = "{Starburst-driven Mass Loss from Dwarf Galaxies: Efficiency and Metal Ejection}",
      journal = {\apj},
         year = 1999,
        month = mar,
       volume = {513},
       number = {1},
        pages = {142-155},
          doi = {10.1086/306832},
archivePrefix = {arXiv},
       eprint = {astro-ph/9801237},
 primaryClass = {astro-ph},
       adsurl = {https://ui.adsabs.harvard.edu/abs/1999ApJ...513..142M}
}

@ARTICLE{Bertone2007,
       author = {{Bertone}, Serena and {De Lucia}, Gabriella and {Thomas}, Peter A.},
        title = "{The recycling of gas and metals in galaxy formation: predictions of a dynamical feedback model}",
      journal = {\mnras},
         year = 2007,
        month = aug,
       volume = {379},
       number = {3},
        pages = {1143-1154},
          doi = {10.1111/j.1365-2966.2007.11997.x},
archivePrefix = {arXiv},
       eprint = {astro-ph/0701407},
 primaryClass = {astro-ph},
       adsurl = {https://ui.adsabs.harvard.edu/abs/2007MNRAS.379.1143B}
}

@ARTICLE{RodriguezRagaCanto2009,
       author = {{Rodr{\'i}guez-Gonz{\'a}lez}, A. and {Raga}, A.~C. and {Cant{\'o}}, J.},
        title = "{Low and high velocity clouds produced by young stellar clusters}",
      journal = {\aap},
         year = 2009,
        month = jul,
       volume = {501},
       number = {2},
        pages = {411-417},
          doi = {10.1051/0004-6361/200911693},
archivePrefix = {arXiv},
       eprint = {0905.1988},
 primaryClass = {astro-ph.GA},
       adsurl = {https://ui.adsabs.harvard.edu/abs/2009A&A...501..411R}
}

@ARTICLE{Rodriguez-gonzalez2011,
       author = {{Rodr{\'i}guez-Gonz{\'a}lez}, A. and {Esquivel}, A. and {Raga}, A.~C. and {Col{\'\i}n}, P.},
        title = "{Mass and metal ejection efficiency in disk galaxies driven by young stellar clusters of nuclear starburst}",
      journal = {\rmxaa},
         year = 2011,
        month = apr,
       volume = {47},
        pages = {113-125},
          doi = {10.48550/arXiv.1102.0234},
archivePrefix = {arXiv},
       eprint = {1102.0234},
 primaryClass = {astro-ph.CO},
       adsurl = {https://ui.adsabs.harvard.edu/abs/2011RMxAA..47..113R}
}

@ARTICLE{Rob-Val2017,
       author = {{Robles-Valdez}, F. and {Rodr{\'\i}guez-Gonz{\'a}lez}, A. and {Hern{\'a}ndez-Mart{\'\i}nez}, L. and {Esquivel}, A.},
        title = "{Metallic Winds in Dwarf Galaxies}",
      journal = {\apj},
         year = 2017,
        month = feb,
       volume = {835},
       number = {2},
          eid = {136},
        pages = {136},
          doi = {10.3847/1538-4357/835/2/136},
archivePrefix = {arXiv},
       eprint = {1612.02787},
 primaryClass = {astro-ph.GA},
       adsurl = {https://ui.adsabs.harvard.edu/abs/2017ApJ...835..136R}
}

@ARTICLE{Krieger2021,
       author = {{Krieger}, N. and {Bolatto}, A.~D. and {Leroy}, A.~K. and {Levy}, R.~C. and {Mills}, E.~A.~C. and {Meier}, D.~S. and {Ott}, J. and {Veilleux}, S. and {Walter}, F. and {Weiss}, A.},
        title = "{VizieR Online Data Catalog: 19 species in 14 super stars clusters in NGC 253 (Krieger+, 2020)}",
      journal = {VizieR Online Data Catalog},
         year = 2021,
        month = oct,
          eid = {J/ApJ/897/176},
        pages = {J/ApJ/897/176},
       adsurl = {https://ui.adsabs.harvard.edu/abs/2021yCat..18970176K}
}

@ARTICLE{Nguyen2022,
       author = {{Nguyen}, Dustin D. and {Thompson}, Todd A.},
        title = "{Galactic Winds and Bubbles from Nuclear Starburst Rings}",
      journal = {\apjl},
         year = 2022,
        month = aug,
       volume = {935},
       number = {2},
          eid = {L24},
        pages = {L24},
          doi = {10.3847/2041-8213/ac86c3},
archivePrefix = {arXiv},
       eprint = {2205.13465},
 primaryClass = {astro-ph.GA},
       adsurl = {https://ui.adsabs.harvard.edu/abs/2022ApJ...935L..24N}
}

@ARTICLE{Brandl_etal_2012A&A...543A..61B,
       author = {{Brandl}, B.~R. and {Mart{\'\i}n-Hern{\'a}ndez}, N.~L. and {Schaerer}, D. and {Rosenberg}, M. and {van der Werf}, P.~P.},
        title = "{High resolution IR observations of the starburst ring in NGC 7552. One ring to rule them all?}",
      journal = {\aap},
         year = 2012,
        month = jul,
       volume = {543},
          eid = {A61},
        pages = {A61},
          doi = {10.1051/0004-6361/201117568},
archivePrefix = {arXiv},
       eprint = {1205.1922},
 primaryClass = {astro-ph.GA},
       adsurl = {https://ui.adsabs.harvard.edu/abs/2012A&A...543A..61B}
}

@ARTICLE{Rekola2005,
       author = {{Rekola}, R. and {Richer}, M.~G. and {McCall}, Marshall L. and {Valtonen}, M.~J. and {Kotilainen}, J.~K. and {Flynn}, Chris},
        title = "{Distance to NGC 253 based on the planetary nebula luminosity function}",
      journal = {\mnras},
         year = 2005,
        month = jul,
       volume = {361},
       number = {1},
        pages = {330-336},
          doi = {10.1111/j.1365-2966.2005.09166.x},
       adsurl = {https://ui.adsabs.harvard.edu/abs/2005MNRAS.361..330R}
}

@ARTICLE{Lopez2023,
       author = {{Lopez}, Sebastian and {Lopez}, Laura A. and {Nguyen}, Dustin D. and {Thompson}, Todd A. and {Mathur}, Smita and {Bolatto}, Alberto D. and {Vulic}, Neven and {Sardone}, Amy},
        title = "{X-Ray Properties of NGC 253's Starburst-driven Outflow}",
      journal = {\apj},
         year = 2023,
        month = jan,
       volume = {942},
       number = {2},
          eid = {108},
        pages = {108},
          doi = {10.3847/1538-4357/aca65e},
archivePrefix = {arXiv},
       eprint = {2209.09260},
 primaryClass = {astro-ph.HE},
       adsurl = {https://ui.adsabs.harvard.edu/abs/2023ApJ...942..108L}
}

@ARTICLE{Meliani2024,
       author = {{Meliani}, Z. and {Cristofari}, P. and {Rodr{\'\i}guez-Gonz{\'a}lez}, A. and {Fichet de Clairfontaine}, G. and {Proust}, E. and {Peretti}, E.},
        title = "{The galactic bubbles of starburst galaxies. The influence of galactic large-scale magnetic fields}",
      journal = {\aap},
         year = 2024,
        month = mar,
       volume = {683},
          eid = {A178},
        pages = {A178},
          doi = {10.1051/0004-6361/202347352},
archivePrefix = {arXiv},
       eprint = {2402.01541},
 primaryClass = {astro-ph.GA},
       adsurl = {https://ui.adsabs.harvard.edu/abs/2024A&A...683A.178M}
}

@ARTICLE{Salpeter1955,
       author = {{Salpeter}, Edwin E.},
        title = "{The Luminosity Function and Stellar Evolution.}",
      journal = {\apj},
         year = 1955,
        month = jan,
       volume = {121},
        pages = {161},
          doi = {10.1086/145971},
       adsurl = {https://ui.adsabs.harvard.edu/abs/1955ApJ...121..161S}
}

@ARTICLE{Leroy2015,
       author = {{Leroy}, Adam K. and {Bolatto}, Alberto D. and {Ostriker}, Eve C. and {Rosolowsky}, Erik and {Walter}, Fabian and {Warren}, Steven R. and {Donovan Meyer}, Jennifer and {Hodge}, Jacqueline and {Meier}, David S. and {Ott}, J{\"u}rgen and {Sandstrom}, Karin and {Schruba}, Andreas and {Veilleux}, Sylvain and {Zwaan}, Martin},
        title = "{ALMA Reveals the Molecular Medium Fueling the Nearest Nuclear Starburst}",
      journal = {\apj},
         year = 2015,
        month = mar,
       volume = {801},
       number = {1},
          eid = {25},
        pages = {25},
          doi = {10.1088/0004-637X/801/1/25},
archivePrefix = {arXiv},
       eprint = {1411.2836},
 primaryClass = {astro-ph.GA},
       adsurl = {https://ui.adsabs.harvard.edu/abs/2015ApJ...801...25L}
}

@ARTICLE{Levy2022,
       author = {{Levy}, Rebecca C. and {Bolatto}, Alberto D. and {Leroy}, Adam K. and {Sormani}, Mattia C. and {Emig}, Kimberly L. and {Gorski}, Mark and {Lenki{\'c}}, Laura and {Mills}, Elisabeth A.~C. and {Tarantino}, Elizabeth and {Teuben}, Peter and {Veilleux}, Sylvain and {Walter}, Fabian},
        title = "{The Morpho-kinematic Architecture of Super Star Clusters in the Center of NGC 253}",
      journal = {\apj},
         year = 2022,
        month = aug,
       volume = {935},
       number = {1},
          eid = {19},
        pages = {19},
          doi = {10.3847/1538-4357/ac7b7a},
archivePrefix = {arXiv},
       eprint = {2206.04700},
 primaryClass = {astro-ph.GA},
       adsurl = {https://ui.adsabs.harvard.edu/abs/2022ApJ...935...19L}
}

@ARTICLE{Veilleux_etal_2005ARA&A..43..769V,
       author = {{Veilleux}, Sylvain and {Cecil}, Gerald and {Bland-Hawthorn}, Joss},
        title = "{Galactic Winds}",
      journal = {\araa},
         year = 2005,
        month = sep,
       volume = {43},
       number = {1},
        pages = {769-826},
          doi = {10.1146/annurev.astro.43.072103.150610},
archivePrefix = {arXiv},
       eprint = {astro-ph/0504435},
 primaryClass = {astro-ph},
       adsurl = {https://ui.adsabs.harvard.edu/abs/2005ARA&A..43..769V}
}

@ARTICLE{ShopbelandBland1998,
       author = {{Shopbell}, P.~L. and {Bland-Hawthorn}, J.},
        title = "{The Asymmetric Wind in M82}",
      journal = {\apj},
         year = 1998,
        month = jan,
       volume = {493},
       number = {1},
        pages = {129-153},
          doi = {10.1086/305108},
archivePrefix = {arXiv},
       eprint = {astro-ph/9708038},
 primaryClass = {astro-ph},
       adsurl = {https://ui.adsabs.harvard.edu/abs/1998ApJ...493..129S}
}

@ARTICLE{RG2007,
       author = {{Rodr{\'\i}guez-Gonz{\'a}lez}, A. and {Cant{\'o}}, J. and {Esquivel}, A. and {Raga}, A.~C. and {Vel{\'a}zquez}, P.~F.},
        title = "{Winds from clusters with non-uniform stellar distributions}",
      journal = {\mnras},
         year = 2007,
        month = sep,
       volume = {380},
       number = {3},
        pages = {1198-1206},
          doi = {10.1111/j.1365-2966.2007.12167.x},
archivePrefix = {arXiv},
       eprint = {0708.1289},
 primaryClass = {astro-ph},
       adsurl = {https://ui.adsabs.harvard.edu/abs/2007MNRAS.380.1198R}
}

@ARTICLE{Canto2000,
       author = {{Cant{\'o}}, J. and {Raga}, A.~C. and {Rodr{\'\i}guez}, L.~F.},
        title = "{The Hot, Diffuse Gas in a Dense Cluster of Massive Stars}",
      journal = {\apj},
         year = 2000,
        month = jun,
       volume = {536},
       number = {2},
        pages = {896-901},
          doi = {10.1086/308983},
       adsurl = {https://ui.adsabs.harvard.edu/abs/2000ApJ...536..896C}
}

@ARTICLE{castellanos2015,
       author = {{Castellanos-Ram{\'\i}rez}, A. and {Rodr{\'\i}guez-Gonz{\'a}lez}, A. and {Esquivel}, A. and {Toledo-Roy}, J.~C. and {Olivares}, J. and {Vel{\'a}zquez}, P.~F.},
        title = "{The soft and hard X-rays thermal emission from star cluster winds with a supernova explosion}",
      journal = {\mnras},
         year = 2015,
        month = jul,
       volume = {450},
       number = {3},
        pages = {2799-2811},
          doi = {10.1093/mnras/stv795},
archivePrefix = {arXiv},
       eprint = {1504.02820},
 primaryClass = {astro-ph.GA},
       adsurl = {https://ui.adsabs.harvard.edu/abs/2015MNRAS.450.2799C}
}

@ARTICLE{Sanders2003,
       author = {{Sanders}, D.~B. and {Mazzarella}, J.~M. and {Kim}, D. -C. and {Surace}, J.~A. and {Soifer}, B.~T.},
        title = "{The IRAS Revised Bright Galaxy Sample}",
      journal = {\aj},
         year = 2003,
        month = oct,
       volume = {126},
       number = {4},
        pages = {1607-1664},
          doi = {10.1086/376841},
archivePrefix = {arXiv},
       eprint = {astro-ph/0306263},
 primaryClass = {astro-ph},
       adsurl = {https://ui.adsabs.harvard.edu/abs/2003AJ....126.1607S}
}

@ARTICLE{Smith1999Apj,
       author = {{Smith}, Denise A. and {Herter}, Terry and {Haynes}, Martha P. and {Neff}, Susan G.},
        title = "{The Luminous Starburst Ring in NGC 7771: Sequential Star Formation?}",
      journal = {\apj},
         year = 1999,
        month = jan,
       volume = {510},
       number = {2},
        pages = {669-686},
          doi = {10.1086/306605},
archivePrefix = {arXiv},
       eprint = {astro-ph/9808331},
 primaryClass = {astro-ph},
       adsurl = {https://ui.adsabs.harvard.edu/abs/1999ApJ...510..669S}
}

@ARTICLE{Zhang2014,
       author = {{Zhang}, Shuinai and {Wang}, Q. Daniel and {Ji}, Li and {Smith}, Randall K. and {Foster}, Adam R. and {Zhou}, Xin},
        title = "{Spectral Modeling of the Charge-exchange X-Ray Emission from M82}",
      journal = {\apj},
         year = 2014,
        month = oct,
       volume = {794},
       number = {1},
          eid = {61},
        pages = {61},
          doi = {10.1088/0004-637X/794/1/61},
archivePrefix = {arXiv},
       eprint = {1408.3207},
 primaryClass = {astro-ph.GA},
       adsurl = {https://ui.adsabs.harvard.edu/abs/2014ApJ...794...61Z}
}

@ARTICLE{smith2014,
       author = {{Smith}, Randall K. and {Foster}, Adam R. and {Edgar}, Richard J. and {Brickhouse}, Nancy S.},
        title = "{Resolving the Origin of the Diffuse Soft X-Ray Background}",
      journal = {\apj},
         year = 2014,
        month = may,
       volume = {787},
       number = {1},
          eid = {77},
        pages = {77},
          doi = {10.1088/0004-637X/787/1/77},
archivePrefix = {arXiv},
       eprint = {1406.2037},
 primaryClass = {astro-ph.HE},
       adsurl = {https://ui.adsabs.harvard.edu/abs/2014ApJ...787...77S}
}

@ARTICLE{Wu2020MNRAS.491.5621W,
       author = {{Wu}, Kinwah and {Li}, Kaye Jiale and {Owen}, Ellis R. and {Ji}, Li and {Zhang}, Shuinai and {Branduardi-Raymont}, Graziella},
        title = "{Charge-exchange emission and cold clumps in multiphase galactic outflows}",
      journal = {\mnras},
         year = 2020,
        month = feb,
       volume = {491},
       number = {4},
        pages = {5621-5635},
          doi = {10.1093/mnras/stz3301},
archivePrefix = {arXiv},
       eprint = {1911.11860},
 primaryClass = {astro-ph.GA},
       adsurl = {https://ui.adsabs.harvard.edu/abs/2020MNRAS.491.5621W}
}

@ARTICLE{Sike2025ApJ,
       author = {{Sike}, Brandon and {Thomas}, Timon and {Ruszkowski}, Mateusz and {Pfrommer}, Christoph and {Weber}, Matthias},
        title = "{Cosmic-Ray-driven Galactic Winds with Resolved Interstellar Medium and Ion-neutral Damping}",
      journal = {\apj},
         year = 2025,
        month = jul,
       volume = {987},
       number = {2},
          eid = {204},
        pages = {204},
          doi = {10.3847/1538-4357/adda3d},
archivePrefix = {arXiv},
       eprint = {2410.06988},
 primaryClass = {astro-ph.GA},
       adsurl = {https://ui.adsabs.harvard.edu/abs/2025ApJ...987..204S}
}

@ARTICLE{Romano2025arXiv,
       author = {{Romano}, Leonard E.~C. and {Owen}, Ellis R. and {Nagamine}, Kentaro},
        title = "{Starburst-Driven Galactic Outflows -- Unveiling the Suppressive Role of Cosmic Ray Halos}",
      journal = {arXiv e-prints},
         year = 2025,
        month = mar,
          eid = {arXiv:2503.13261},
        pages = {arXiv:2503.13261},
          doi = {10.48550/arXiv.2503.13261},
archivePrefix = {arXiv},
       eprint = {2503.13261},
 primaryClass = {astro-ph.GA},
       adsurl = {https://ui.adsabs.harvard.edu/abs/2025arXiv250313261R}
}

@ARTICLE{SandersFabian2011MNRAS,
       author = {{Sanders}, J.~S. and {Fabian}, A.~C.},
        title = "{Revealing O VII from stacked X-ray grating spectra of clusters, groups and elliptical galaxies}",
      journal = {\mnras},
         year = 2011,
        month = mar,
       volume = {412},
       number = {1},
        pages = {L35-L39},
          doi = {10.1111/j.1745-3933.2010.01000.x},
archivePrefix = {arXiv},
       eprint = {1012.0235},
 primaryClass = {astro-ph.CO},
       adsurl = {https://ui.adsabs.harvard.edu/abs/2011MNRAS.412L..35S}
}

@INPROCEEDINGS{Grevesse1989AIPC,
       author = {{Grevesse}, Nicolas and {Anders}, Edward},
        title = "{Solar-system abundances of the elements: A new table}",
    booktitle = {Cosmic Abundances of Matter},
         year = 1989,
       editor = {{Waddington}, C. Jake},
       series = {American Institute of Physics Conference Series},
       volume = {183},
        month = jan,
    publisher = {AIP},
        pages = {1-8},
          doi = {10.1063/1.38013},
       adsurl = {https://ui.adsabs.harvard.edu/abs/1989AIPC..183....1G}
}

@ARTICLE{Beck1994,
       author = {{Beck}, R. and {Carilli}, C.~L. and {Holdaway}, M.~A. and {Klein}, U.},
        title = "{Multifrequency observations of the radio continuum emission from NGC 253. I. Magnetic fields and rotation measures in the bar and halo.}",
      journal = {\aap},
         year = 1994,
        month = dec,
       volume = {292},
        pages = {409-424},
       adsurl = {https://ui.adsabs.harvard.edu/abs/1994A&A...292..409B}
}
\appendix
\section{Emission}
\label{sec:emis}

\subsection{Soft X-ray emission}

To calculate the X-ray emission for our models, we used the coefficients of CHANTI and the integration obtained in \cite{castellanos2015}, in the range of 0.3 to 10~keV, considering a solar metallicity. We  fit a polynomial function for the coefficients.

\begin{figure}[!h]
    \centering
    \includegraphics[width=\columnwidth]{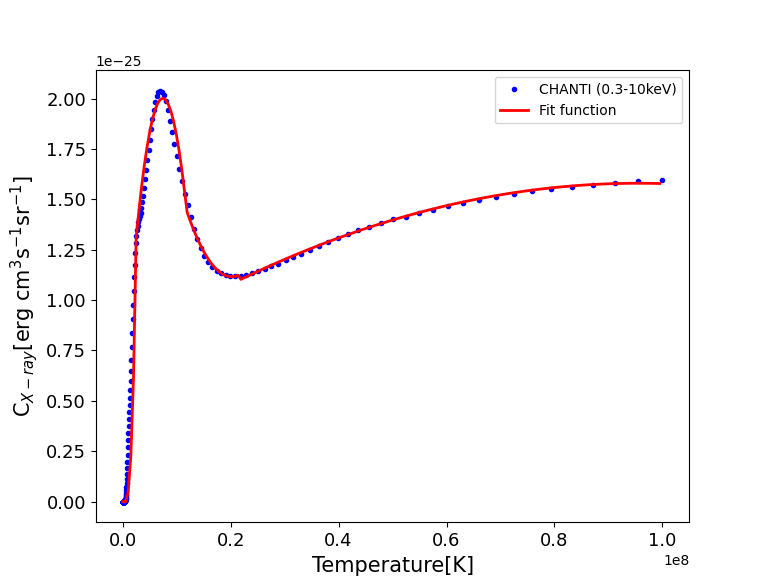}
    \caption[]{Fit function for the X-ray emission coefficient (0.3 to 10~keV) compared with the CHANTI data. }
    \label{fig:Xfit}
\end{figure}

\subsection{H$\alpha$ emission}

For the H$\alpha$, we took  the temperature and density of each cell of the simulation and calculated the collision and recombination coefficients as in Raga et al. (2015),
 \begin{equation}
\epsilon_r(\rm{H}\alpha)=\frac{a}{T^{0.568}+b\;T^{1.5}} ,
 \end{equation}
 and
 \begin{equation}
     \epsilon_c (\rm{H}\alpha)=\frac{c}{T^{0.5}}e^{-d/T}\times \left( 1+ \frac{f}{1+g/T} \right),
 \end{equation}
where a=4$\times$10$^{-23}$, b=3.85$\times$10$^{-5}$ , c=$3.57\times 10^{-17}$, d$=140360$, f$=7.8$
 y g$= 5\times 10^5$, and  T is the electron temperature.

Then, the  H$\alpha$ coefficient is given by
\begin{equation}
    j_{H\alpha}=\frac{1}{4 \pi} \left [\chi_{ \rm HII}^2 n_H^2\left(\chi_{ \rm HII}\; \epsilon_r({\rm H}\alpha)+(1-\chi_{ \rm HII}) \, \epsilon_c ({\rm H}\alpha_c)\right)\right],
\end{equation}
where $\chi_{ \rm HII}$ is the ionisation fraction, considering a coronal equilibrium given by\\

\begin{equation}
    \chi_{ \rm HII}=\frac{1}{1+\alpha(T)/c(T)}
\end{equation}

With  \(\alpha(T)=3.69\times 10^{-10}T^{-0.79}\) and $c(T)=5.83 \times 10^{-11} T^{0.5}e^{157800/T}$.

\end{document}